\documentclass[preprint,12pt]{elsarticle}

\usepackage{booktabs}
\usepackage{blindtext}

\usepackage{amsmath}
\usepackage{amssymb}
\usepackage{mathtools}
\usepackage{array}
\usepackage{mathrsfs}
\usepackage{mathalpha}
\usepackage{cite} 
\usepackage{longtable} 
\usepackage[font=small,skip=2pt]{caption} 
\usepackage{float} 

\numberwithin{equation}{section}

\journal{Information Sciences}

\begin{document}

\begin{frontmatter}



\title{Information Science Principles of Machine Learning:\\ A Causal Chain Meta-Framework Based on Formalized Information Mapping}


\author{Jianfeng Xu} 

\affiliation{organization={Koguan School of Law, China Institute for Smart Justice, School of Computer Science, Shanghai Jiao Tong University},
            city={Shanghai},
            postcode={200030},
            country={China}}

\begin{abstract}
This paper addresses the current lack of a unified formal framework in machine learning theory, as well as the absence of robust theoretical foundations for interpretability and ethical safety assurance. We first construct a formal information model that employs sets of well-formed formulas (WFFs) to explicitly define the ontological states and the carrier mappings of core ML components. By introducing learnable and processable predicates, as well as learning and processing functions, we analyze the logical inference and constraint rules underlying causal chains in models, thereby establishing the Machine Learning Theory Meta-Framework (MLT-MF). Building upon this framework, we propose universal definitions for model interpretability and ethical safety, and rigorously prove four key theorems: (i) the equivalence between model interpretability and information existence; (ii) a constructive formulation of ethical safety assurance; and (iii–iv) two upper bounds on total variation distance (TVD).This work overcomes the limitations of previous fragmented approaches, providing a unified theoretical foundation from an information science perspective to systematically address the critical challenges currently facing machine learning.
\end{abstract}



\begin{keyword}
machine learning, objective information theory (OIT), formal logic systems, interpretability, ethical safety assurance, error bound estimation
\end{keyword}

\end{frontmatter}


\section{Introduction}

Machine learning (ML) has progressed rapidly, yet several foundational gaps persist. The field still lacks a unified logical framework that spans the full lifecycle from prior specification and data-driven learning to deployment and online adaptation. Existing theory remains fragmented across training dynamics, initialization, test-time generalization, and deployment, which prevents a coherent account of knowledge evolution across stages \citep{vapnik1999overview,alquier2024user,scholkopf2021toward,garcez2023neurosymbolic,marra2024statistical}. Model performance is also hard to predict, which inflates development costs in data, computation, energy, and human effort. In the absence of explicit causal semantics, structural analysis, and stage-wise error metrics, attainable performance levels and error upper bounds are difficult to estimate, and information coverage gaps or failure modes due to model selection, measurement, and data quality are hard to anticipate \citep{achiam2023gpt4,touvron2023llama,chowdhery2023palm,gunasekar2023textbooks}. Ethical safety lacks a settled mathematical definition and verification protocol. Current research is dominated by principles and heterogeneous guidelines, leaving uncertainty about standards and assurance processes \citep{russell2015research,burr2023ethical,tilala2024ethical,hawkins2021guidance,pant2024ethics}. Interpretability similarly lacks a rigorous, domain-agnostic foundation. Without a unified semantic framework for causal inference and constraint logic, the data-to-model pipeline is treated as a black box, and general, testable definitions and criteria remain elusive despite numerous approaches \citep{lipton2018mythos,glanois2024survey,zhang2024robustness,rudin2019stop,miller2019explanation}.

Information science offers a route to unification \citep{shwartz2017opening,xu2022information}, but its subfields—information theory, control, signal processing, statistical learning, and computation—evolved largely in isolation and lack a common semantics. Objective Information Theory (OIT) addresses this by axiomatizing information as an enabling mapping from ontological states to carrier states. Intrinsic meaning is captured by ontological states and objective expression by carrier states, and the mapping has both mathematical and physical–causal attributes \citep{xu2014objective,xu2024research}. Prior work shows that OIT’s general metrics can improve data evaluation and training efficiency in machine learning \citep{xu2025general}. We complete a remaining gap by formalizing state as an interpretation of well-formed formulas (WFFs) in a higher-order logical system, and by clarifying that enabling mappings are surjective mathematical mappings that are also physically realizable causal relations. We further introduce noisy information and information chains, which support analysis of multi-stage pipelines and their causal dependencies.

Building on this foundation, we develop the Machine Learning Theory Meta-Framework (MLT-MF), a unified information–logical semantics for the ML lifecycle. Our main contributions are:

•	Unified lifecycle semantics. We formalize five canonical stages—model construction, training information, learning and training, problem input, and processing/output with online learning—via state objects and enabling mappings that compose into information chains.

•	Logical definitions with verifiable criteria. We give domain-general, testable definitions of interpretability and ethical safety within the same semantics. We prove a necessary and sufficient condition linking interpretability to information existence along the chain and construct a maximal safe-output (SafeMax) mechanism validated via a hypergraph independent-set formulation.

•	Error control with observable quantities. We derive explicit total variation distance (TVD) upper bounds in two regimes: conservative prediction in regions not covered by training under noise-free input, and arbitrary prediction under noisy input. The bounds are driven by measurable masses—overlap, uncovered, and noise (loss/addition)—and align with engineering indicators such as coverage and Kullback–Leibler (KL) divergence \citep{vapnik1999overview,bartlett2005local,xu2017information,wang2024theoretical,huang2025convergence,goodfellow2016deep,pinsker1964information,rodriguez2021tighter,devroye2013probabilistic}.

•	Practical instantiation. On a simple feedforward network, the framework yields multi-modal interpretability through mathematical, visual, and partition-based explanations, guarantees ethical safety via SafeMax, and performance estimates via TVD bounds, illustrating deployable assurance across stages.

The remainder of the paper is organized as follows. Section 2 reviews related work on learning theory, interpretability, ethical safety, and error estimation, and situates OIT as a unifying lens. Section 3 presents theoretical basis: definition of information, noisy information, information chains, and the finite-automaton view. Section 4 introduces MLT-MF, defining stages, predicates, and functions, and specifying lifecycle constraints and inference rules with a minimal pipeline instantiation. Section 5 provides formal definitions and theorems for interpretability and ethical safety, the SafeMax mechanism, and the TVD bounds. Section 6 reports the neural-network case study and observable metrics. Section 7 discusses limitations and future directions.

\section{Related Work}

\subsection{Conceptualizations of Machine Learning}

Classical statistical learning theory formalizes learning as risk minimization over function classes under unknown data-generating distributions, with foundational tools such as VC theory, growth functions, and distribution-independent bounds \citep{vapnik1999overview}. Parameterized predictors, loss-based risk, and empirical risk minimization provide the standard optimization lens \citep{alquier2024user}. Recent perspectives emphasize causal structure beyond observed distributions, seeking representations that expose latent causal factors to support robust transfer and planning across environments \citep{scholkopf2021toward}. Neuro-symbolic approaches pursue integration of distributed representations with symbolic abstraction and compositional reasoning, aiming to combine efficient learning with discrete inference \citep{garcez2023neurosymbolic,marra2024statistical}. These lines of work clarify individual components of the learning pipeline, yet they do not supply a unified semantic and causal framework that spans model construction, training, input, inference, and deployment.

\subsection{Interpretability}

Despite extensive interest, interpretability lacks a precise and generally accepted definition and evaluation protocol. Prior analyses highlight ambiguity in terminology and goals, and suggest that the practical objective is to extract useful information from models in ways that are audience- and task-sensitive \citep{lipton2018mythos}. In reinforcement learning, interpretability depends on interpretable inputs, models, and decisions, and on notions such as simulatability, decomposability, and algorithmic transparency \citep{glanois2024survey}. Mechanistic studies relate training mechanisms to properties such as smoothness and gradient similarity to explain transferability and adversarial behavior, illustrating how dissecting key factors in training can yield explanatory leverage \citep{zhang2024robustness}. Overall, current methods are diverse and often model- or context-specific, and they lack a unifying logical semantics that ties inputs, internal states, and outputs to causal explanations across the lifecycle \citep{lipton2018mythos,glanois2024survey,zhang2024robustness,rudin2019stop,miller2019explanation}.

\subsection{Ethical Safety Assurance}

As AI systems permeate high-stakes domains, ethical safety has drawn sustained attention in law, policy, and engineering. Existing work emphasizes assurance processes grounded in structured arguments that justify normative claims with evidence across the system lifecycle \citep{russell2015research,burr2023ethical}. In healthcare, recommended practices include data minimization, diversity and augmentation, algorithmic fairness, auditing, interpretability, and standards-based oversight, typically framed as a toolbox of technical and procedural controls \citep{tilala2024ethical}. Safety guidance for autonomous and software-intensive systems links ML safety to software and system safety via requirement derivation and lifecycle-wide assurance activities \citep{hawkins2021guidance}. Practitioner studies report gaps in clear moral knowledge and the ambiguity of principles, which complicate operationalization and adoption \citep{pant2024ethics}. What remains missing is a mathematically grounded definition of ethical safety with constructible guarantees and verifiable criteria that integrate with ML pipelines.

\subsection{Model Error Estimation}

Foundational results establish consistency and finite-sample bounds for empirical and structural risk minimization through VC dimension and related complexity measures, including conditions for one- and two-sided uniform convergence and margin-based generalization \citep{vapnik1999overview}. Local Rademacher complexity sharpens bounds by tailoring complexity to low-error regions and yields optimal or near-optimal rates via fixed-point analyses under variance control \citep{bartlett2005local}. Information-theoretic approaches bound generalization via mutual information between data and hypotheses, introduce information stability, and analyze Gibbs and noisy ERM with mutual-information regularization \citep{xu2017information}. In meta-reinforcement learning, generalization across tasks and convergence under nonconvex optimization are characterized using empirical process theory and concentration tools, with guidance for capacity and learning-rate choices \citep{wang2024theoretical}. For generative modeling, recent analyses provide TVD bounds for deterministic samplers via ODE perspectives and establish discrete-time error under Runge–Kutta schemes \citep{huang2025convergence}. These advances quantify error in specific settings but do not explicitly model causal semantics across multi-stage information flows, nor do they integrate noise structure, coverage geometry, and uncovered-region strategies into a single computable bound that can guide data auditing and prediction strategy design.

\subsection{Toward a Unified Information-Theoretic and Logical Perspective}

Information science offers a unifying lens, yet its subfields have developed largely in silos \citep{shwartz2017opening,xu2022information}. OIT axiomatizes information as an enabling mapping from ontological to carrier states with dual mathematical and physical–causal attributes, and provides general metrics that have shown value for data evaluation in ML \citep{xu2014objective,xu2024research,xu2025general}. By formalizing states as interpretations of WFFs in a higher-order logical system and introducing notions of noisy information and information chains, OIT supports analysis of multi-stage pipelines and causal dependence structures \citep{xu2014objective,xu2024research,qiu2025research,wiener2019cybernetics}. The present work builds on this foundation to offer a single formal semantics for lifecycle reasoning about interpretability, ethical safety, and computable TVD-based error control.

\section{Theoretical Basis}

Modern ML systems are information systems and, in the mathematical sense, finite automata. We ground our meta-framework in a formal account of information and state that supports causal analysis across stages.

\subsection{A Unified Information Model of Semantic Intension, Statistical Characteristics, and Physical Realization}

\subsubsection{Information as Enabling Mapping}
\citep{xu2024research} based on four fundamental postulates, proposes that information is an enabling mapping from the state of an ontology to the state of a carrier, but does not further elaborate on the notion of “enabling mapping”. Here, we proceed to further clarify the concept.

\textbf{Definition 1 (Enabling Mapping)} Let  $X $ and  $Y $ be two sets, and let  $f $ be a mapping. If for any  $x \in X $, there exists a  $y \in Y $, and for any  $y \in Y $, there exists at least one  $x \in X $ such that  $y = f(x) $, and furthermore, the existence of  $y $ is physically realized only because of  $x $, then  $f $ is called an enabling mapping from  $X $ to  $Y $. This is denoted as:
\begin{equation}
f:X \rightrightarrows Y
\end{equation}

In (3.1), the symbols  $\rightrightarrows $ is used to indicate enabling mapping, which differs from the standard notation  $\rightarrow $ and used for ordinary mapping. This distinction signifies that  $X $ not only establishes a surjective mapping with  $Y $ in the mathematical sense, but also forms a physical causal enabling relationship between the two. Moreover, its realization requires suitable physical conditions and energy (Fig. 1(a)).

\begin{figure*}[ht]
\centering
\includegraphics[width=0.9\textwidth]{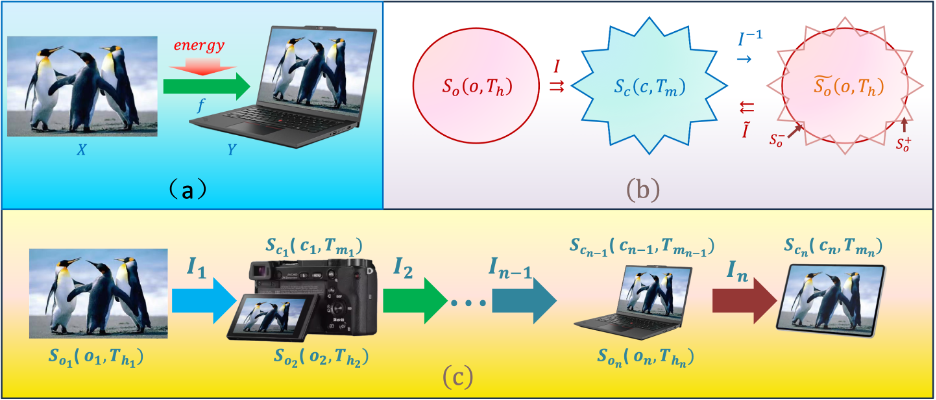}
\caption{Schematic Diagram of Enabling Mapping, Noisy Information, and Information Chain.}
\label{fig:1}
\end{figure*}

Based on Appendix A’s higher-order formal system $\mathcal{L} $ and the result that virtually any state set can be represented as the interpretation of a set of WFFs in $\mathcal{L} $ \citep{qiu2025research}, we obtain a more robust theoretical foundation for defining information:

\textbf{Definition 2 (Definition of Information)} Let $\mathbb{O} $ denote the set of objective entities,  $\mathbb{S} $ the set of subjective entities, and  $\mathbb{T} $ the set of times, with elements of  $\mathbb{O} $,  $\mathbb{S} $, and  $\mathbb{T} $ partitioned as needed depending on the domain. The ontology $o\mathbb{\subseteq O} \cup \mathbb{S} $, time of occurrence  $T_{h}\mathbb{\subseteq T} $, ontological state set  $S_{o}\left( o,T_{h} \right) $, carrier  $c\mathbb{\subseteq O} $, reflection time  $T_{m}\mathbb{\subseteq T} $, and carrier reflection set  $S_{c}\left( c,T_{m} \right) $ are all non-empty sets. Then, information  $I $ is the enabling mapping from  $S_{o}\left( o,T_{h} \right) $ to  $S_{c}\left( c,T_{m} \right) $, succinctly denoted as $I:S_o( o,T_h ) \rightrightarrows S_{c}( c,T_{m})$ or $I = \left\langle o,T_{h},S_{o},c,T_{m},S_{c} \right\rangle $. 

The set of all information  $I $ is called the information space, denoted by  $\mathfrak{T} $, which, together with matter and energy, forms the three fundamental constituents of the objective world \citep{wiener2019cybernetics}. Definition 2 subsumes Shannon’s model: source events and channel codes are states, and source-to-channel is an enabling mapping; OIT thus generalizes classical information theory \citep{shannon1948mathematical}.

\subsubsection{Ideal Information}

\textbf{Definition 3 (Semantic Inclusion)}
Let $S$ be a foundational semantic system. For a logical system $L$, if there exists a semantic mapping $\mathrm{Sem}_L$ from $L$ into $S$ such that, for every symbol $x$ in $L$, $x$ is synonymous with $\mathrm{Sem}_L(x)$, then $S$ is said to \emph{semantically include} $L$. If $L_1$ and $L_2$ are both semantically included by $S$, then $L_1$ and $L_2$ are said to be \emph{synonymous relative to $S$}.

\textbf{Definition 4 (Synonymous States)}
Let $S_{x_1}(x_1,t_1)$ and $S_{x_2}(x_2,t_2)$ be sets of states expressible in logical systems $L_1$ and $L_2$, respectively. If $L_1$ and $L_2$ are synonymous relative to $S$, and for every $\varphi_1 \in S_{x_1}(x_1,t_1)$ there exists $\varphi_2 \in S_{x_2}(x_2,t_2)$, and for every $\varphi_2 \in S_{x_2}(x_2,t_2)$ there exists $\varphi_1 \in S_{x_1}(x_1,t_1)$, such that $\mathrm{Sem}_{L_1}(\varphi_1) = \mathrm{Sem}_{L_2}(\varphi_2)$, then $S_{x_1}(x_1,t_1)$ and $S_{x_2}(x_2,t_2)$ are said to be \emph{synonymous relative to $S$}, or simply \emph{synonymous}.

\textbf{Definition 5 (Ideal Information)}
Consider information $I: S_o(o,T_h) \rightrightarrows S_c(c,T_m)$. If $S_o(o,T_h)$ and $S_c(c,T_m)$ are synonymous relative to the foundational semantic system $S$, then $I$ is called \emph{ideal information with respect to $S$}; when no confusion arises, we may simply call $I$ \emph{ideal information}.

Since the ontological state $S_o(o,T_h)$ and the carrier state $S_c(c,T_m)$ of ideal information $I$ are synonymous, either $S_o(o,T_h)$ or $S_c(c,T_m)$ can be taken to represent $I$.

\subsubsection{The Semantic Intension of Information}

\textbf{Definition 6 (Explicit Assertions and Explicit Information Amount)}
For the ontological state $S_o(o,T_h)$ of ideal information $I$, write $S_o = \{\varphi \mid \varphi \text{ is a WFF in logical system } L \text{ under an interpretation over } o \text{ and } T_h\}$. Each $\varphi$ is an \emph{explicit assertion} of $I$; hence $S_o$ is the set of explicit assertions of $I$, and $|S_o|$ is the \emph{explicit information amount} of $I$.

\textbf{Definition 7 (Logical Consequences and Implicational Information Amount)}
For the ontological state $S_o$ of ideal information $I$, the logical closure $\mathrm{Cn}(S_o)$ contains all logical consequences of $I$, and $|\mathrm{Cn}(S_o)|$ is the \emph{implicational information amount} of $I$.

\textbf{Definition 8 (Atomic Information and Primitive Information Amount)}
For the ontological state $S_o$ of ideal information $I$, suppose $\varphi$ cannot be derived from the other WFFs in $S_o$; then $\varphi$ is called an \emph{atomic state} of $S_o$, and $I(\varphi)$ is an \emph{atomic piece of information} in $I$. Let $\mathrm{Atom}(S_o) = \{\varphi \mid \varphi \text{ is an atomic state in } S_o\}$. Then $|\mathrm{Atom}(S_o)|$ is the \emph{primitive information amount} of $I$.

Explicit assertions, logical consequences, and ontic atomic states all belong to the semantic intension of information.

\textbf{Definition 9 (Logical Depth)}
For the ontological state $S_o$ of ideal information $I$, let $\varphi \in S_o$ or $\varphi \in \mathrm{Cn}(S_o)$. If deriving $\varphi$ from the atomic states of $S_o$ requires $i$ ($i=0,1,\dotsm$) recursive layers, then the \emph{logical depth} of $\varphi$ in $S_o$ is $i$, denoted $\mathrm{Ld}(\varphi)$. The logical depths of $S_o$ and $\mathrm{Cn}(S_o)$ are defined by $\mathrm{Ld}(S_o) = \max\{\mathrm{Ld}(\varphi) \mid \varphi \in S_o\}$ and $\mathrm{Ld}(\mathrm{Cn}(S_o)) = \max\{\mathrm{Ld}(\varphi) \mid \varphi \in \mathrm{Cn}(S_o)\}$, respectively.

\subsubsection{The Extensional Characteristics of Information}

\textbf{Definition 10 (Extensional Characteristics of Information)}
The \emph{extensional characteristics} of information $I: S_o(o,T_h) \rightrightarrows S_c(c,T_m)$ in a logical system $L$ are the states $S(S_o \cup S_c)$ expressible in $L$ by $S_o(o,T_h)$ and $S_c(c,T_m)$, denoted $\mathrm{Ext}(I)$.

Statistical characteristics are an important component of extensional characteristics. For ideal information $I$, the explicit information amount $|S_o|$, the implicational information amount $|\mathrm{Cn}(S_o)|$, the primitive information amount $|\mathrm{Atom}(S_o)|$, and the logical depths $\mathrm{Ld}(S_o)$ and $\mathrm{Ld}(\mathrm{Cn}(S_o))$ are all statistical characteristics and belong to $\mathrm{Ext}(I)$.

\textbf{Definition 11 (Information Entropy)}
For the ontological state $S_o = \{\varphi_i \mid i=1,\dotsm,|S_o|\}$ of ideal information $I$, suppose the probability of each $\varphi_i$ is $p_i$ ($i=1,\dotsm,|S_o|$). Then $H(I) = \sum_{i=1}^{|S_o|} p_i$ is the \emph{information entropy} of $I$. It also belongs to the statistical characteristics of $\mathrm{Ext}(I)$.

\subsubsection{Physical Realization of Information}

The physical realization of information is an enabling mapping. For information $I: S_o(o,T_h) \rightrightarrows S_c(c,T_m)$, practical circumstances---due to constraints such as channel conditions, environment, energy, and time---make it difficult to ensure that $S_o(o,T_h)$ and $S_c(c,T_m)$ are synonymous relative to the foundational semantic system $S$.

\textbf{Definition 12 (Noisy Information)}
For information $I: S_o(o,T_h) \rightrightarrows S_c(c,T_m)$, it is not difficult to show that there always exists a set $\widetilde{S_o}(o,T_h) = S_o(o,T_h) \setminus S_o^-(o,T_h) \cup S_o^+(o,T_h)$ such that $\widetilde{S_o}(o,T_h)$ and $S_c(c,T_m)$ are synonymous relative to $S$. Here $S_o^-$ is called the \emph{noise-loss state set}, with $S_o^- \subseteq S_o$; $S_o^+$ is called the \emph{noise-addition state set}, with $S_o \cap S_o^+ = \emptyset$. In this case, $S_c(c,T_m)$ can also be viewed as the result of an enabling mapping from $\widetilde{S_o}(o,T_h)$, namely $\widetilde{I}: \widetilde{S_o}(o,T_h) \rightrightarrows S_c(c,T_m)$, which is called the \emph{noisy information} of $I$.

Fig. 1(b) depicts the structural characteristics of noisy information. In Shannon’s terms, noise transforms the source state via deletion and addition before transmission; the resulting channel output realizes the noisy carrier state \citep{shannon1948mathematical}. This structural decomposition will support our TVD bounds by isolating loss and addition masses.

\subsection{Information Chain}

\citep{xu2014objective} uses composition to discuss the transitivity of information. Specifically, if $I_o:S_o( o,T_h ) \rightrightarrows S_{c_1}( c_1,T_{m_1})$ and $I_1:S_{c_1 } (c_1,T_{m_1 }) \rightrightarrows S_{c_2 } (c_2,T_{m_2 })$ are two information instances, then $I_2:S_o (o,T_h) \rightrightarrows S_{c_2 } (c_2,T_{m_2 })$ is also an information instance, interpreted as $I_0$ being transmitted from carrier $c_1$ to carrier $c_2$. This yields the notion of an information chain.

Let $\left\{ I_i: S_{o_i } (o_i,T_{h_i }) \rightrightarrows S_{c_i } (c_i,T_{m_i }) \middle| i = 1,\cdots,n \right\} $
be a sequence of information mappings, where for any  $i > 1 $, it holds that  $o_{i} = c_{i - 1} $, $T_{h_{i}} = T_{m_{i - 1}} $, and  $S_{o_{i}}\left( o_{i},T_{h_{i}} \right) = S_{c_{i - 1}}(c_{i - 1},T_{m_{i - 1}}) $. In this case, we say that these mappings constitute an information chain. Moreover, for any 1 $\leq i \leq j \leq n $, $I_{ij}: S_{o_i } (o_i,T_{h_i }) \rightrightarrows S_{c_j } (c_j,T_{m_j }) $ also qualifies as information according to Definition 2.

Transmission constitutes a fundamental mode of information flow and application. In many cases, information undergoes more complex flows through chain-like transmission processes. Fig. 1(c) intuitively illustrates such a process. For any pair of states within this process, the latter can be regarded as the result of the enabling mapping from the former, and this enabling mapping can be used to reasonably explain the causation of the latter state. This provides a theoretical foundation for the interpretability of the model.

\subsection{Finite Automata as Formal State Systems}

Any practical computing or ML system can be formalized as a finite automaton with state, input, output, and transition functions over discrete times (Appendix B). We represent its behavior by a state set of WFFs that captures inputs, outputs, and transitions, together with predicates and functions for assignability, learnability, and processability. This provides the operational substrate for our lifecycle formalization in Section 4 and for the logical definitions and theorems in Section 5.

\section{Meta-Framework for Machine Learning Theory (MLT-MF)}

The core of ML is the artificial intelligence model, which can be regarded as a finite automaton as well as an information system. The processes of operation, learning, testing, and application for such a model all fundamentally depend on information within the system. Therefore, by employing the concepts of information model and finite automaton state, the process of ML and its applications can be unified into five primary steps of information realization and functioning. Fig. 2 presents the chain-structured architecture of causal information progression across the five stages. The subsequent subsections provide formal expressions of the information states for each stage. Undoubtedly, these pieces of information are the fundamental basis for explaining model behavior and outcomes. They transcend specific algorithms or models and provide a new theoretical framework—MLT-MF—for addressing a range of critical issues such as ML interpretability.

\begin{figure*}[ht]
\centering
\includegraphics[width=0.75\textwidth]{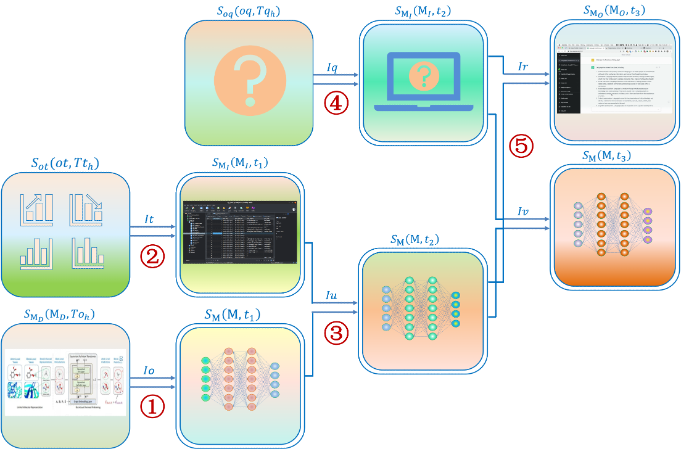}
\caption{Causal Chains of Multiple Information Mappings in Machine Learning.}
\label{fig:2}
\end{figure*}

\subsection{Construction of Initial Information}

For the first stage in Fig. 2, suppose the model under study is denoted as $M $, and let $T = \left\{ t_{i} \middle| i = 1,\ldots,n \right\} $ be the set of time points at which $M $, as a finite automaton, undergoes assignment, input, output, and state transitions in increasing order. According to the perspective of OIT, the creation of $M $ is essentially the realization of information:
\begin{equation}
Io:S_{M_D } (M_D,To_h) \rightrightarrows S_M (M,t_1)
\end{equation}
where the ontology  $M_{D} $ of  $Io $ is the designer's conceptualization of  $M $,  ${To}_{h} $  is the time at which the designer conceives  $M $, and  $S_{M_{D}}\left( M_{D},{To}_{h} \right) $ is the initial finite automaton state that must be realized at this conceptual stage. The carrier  $M $ of  $Io $   is the model itself; the reflection
time  $t_{1} $  is when  $M $ is constructed according to  $S_{M_{D}}\left( M_{D},{To}_{h} \right) $; and  $S_{M}\left( M,t_{1} \right) $ is the formalization of the actual physical state of  $M $ upon its completion. Thus,  $max( {To}_{h} ) < max(t_{1}) $.

According to Lemma B.2, let the binary function  $\Phi^{2}(X,t) $ represent the set of structures, parameters, and algorithms of model $X $ at time  $t $. The formula
\begin{equation}
{\varphi o}_{1} = {State}^{3}\left( M_{D},t_{1},\Phi^{2}(M_{D},{To}_{h}) \right)
\end{equation}
denotes that, at time  $t_{1} $, the initial structure, parameters, and algorithm set  $\Phi^{2}(M_{D},{To}_{h}) $ as envisioned by the designer are assigned to  $M_{D} $.
\begin{equation}
\begin{aligned}
\varphi o_2 = & \mathit{State}^{3}(M_{D}, t_{1}, \Phi^{2}(M_{D}, To_{h})) \land \\
& \mathit{Input}^{3}(M_{D}, t_{1}, S_{M_{I}}(M_{I}, t_{1})) \land \\
& \mathit{Learnable}^{2}(M_{D}, S_{M_{I}}(M_{I}, t_{1})) \rightarrow \exists f_{\mathit{learn}}^{2}(M_{D}, S_{M_{I}}(M_{I}, t_{1})) \\
& \left( \mathit{Eq}^{2}(\Phi^{2}(M_{D}, t_{2}), f_{\mathit{learn}}^{2}(M_{D}, S_{M_{I}}(M_{I}, t_{1}))) \right) \land \\
& \mathit{State}^{3}(M_{D}, t_{2}, \Phi^{2}(M_{D}, To_{h}))
\end{aligned}
\end{equation}
expresses that, at time  $t_{1} $, when the structure, parameters, and algorithm set of  $M_{D} $  is  $\Phi^{2}(M_{D},{To}_{h}) $, and the input carrier state  $S_{M_{I}}\left( M_{I},t_{1} \right) $ is provided and satisfies the learnability condition determined by the binary predicate  ${Learnable}^{2}(M_{D},S_{M_{I}}\left( M_{I},t_{1} \right)) $,
 $M_{D} $  can optimize its internal structures, parameters, and algorithms to  $\Phi^{2}(M_{D},t_{2}) $ through the learning binary function  $f_{learn}^{2}(M_{D},S_{M_{I}}\left( M_{I},t_{1} \right)) $, assigning this to  $M_{D} $. This demonstrates the model\textquotesingle s capacity for learning and optimization, where  $M_{I} $  denotes the input device of  $M $.

\begin{equation}
\begin{aligned}
\varphi o_3 = & \mathit{State}^{3}(M_{D}, t_{1}, \Phi^{2}(M_{D}, To_{h})) \land \\
& \mathit{Input}^{3}(M_{D}, t_{1}, S_{M_{I}}(M_{I}, t_{1})) \land \\
& \mathit{Processable}^{2}(M_{D}, S_{M_{I}}(M_{I}, t_{1})) \rightarrow \exists f_{\mathit{process}}^{2}(M_{D}, S_{M_{I}}(M_{I}, t_{1})) \\
& \left( \mathit{Eq}^{2}(S_{M_{O}}(M_{O}, t_{2}), f_{\mathit{process}}^{2}(M_{D}, S_{M_{I}}(M_{I}, t_{1}))) \right) \land \\
& \mathit{Output}^{3}(M_{D}, t_{2}, S_{M_{O}}(M_{O}, t_{2}))
\end{aligned}
\end{equation}
{\sloppy\emergencystretch=1em
indicates that, at time  $t_{1} $, if  $S_{M_{I}}\left( M_{I},t_{1} \right) $ is input into $M_{D} $  and meets the processability condition specified by the binary predicate ${Processable}^{2}(M,S_{M_{I}}\left( M_{I},t_{1} \right)) $, then  $M_{D} $  can process  $S_{M_{I}}\left( M_{I},t_{1} \right) $ via the binary processing function  $f_{process}^{2}(M,S_{M_{I}}\left( M_{I},t_{1} \right)) $ to yield the output carrier state  $S_{M_{O}}\left( M_{O},t_{2} \right) $, which is then output at time  $t_{2} $. This demonstrates the model's ability to process and output information, where  $M_{O} $ denotes the output device of  $M $. When output information must not violate relevant ethical guidelines, $f_{process}^2$ must implement the corresponding safety assurance functions.
\par}

Thus,
\begin{equation}
S_{M_{D}}\left( M_{D},{To}_{h} \right) = \left\{ {\varphi o}_{1},{\varphi o}_{2},{\varphi o}_{3} \right\}
\end{equation}
constitutes the initial state of  $M $ as conceived by the designer.

{\sloppy\emergencystretch=1em
It should be noted that, since  $M_{D} $,  $\Phi^{2}(M_{D},{To}_{h}) $, and ${State}^{3}\left( M_{D},t_{1},\Phi^{2}(M_{D},{To}_{h}) \right) $ all represent sets, ${\varphi o}_{1},{\varphi o}_{2},{\varphi o}_{3} $  may be considered as three WFFs, or---when decomposed with respect to the elements of each set---as three sets of more granular WFFs. Regardless of the situation, both cases satisfy Definition B.1 concerning finite automaton states. Therefore, subsequent discussions will encompass both cases without further elaboration.
\par}

 $S_{M_{D}}\left( M_{D},{To}_{h} \right) $ may represent the designer's idea, a design draft on paper, or an electronic document. Through the information  $Io $, these are enabled and mapped into the physically operational state of the model  $M $:
\begin{equation}
S_{M}\left( M,t_{1} \right) = \left\{ {\psi o}_{1},{\psi o}_{2},{\psi o}_{3} \right\}
\end{equation}
where  ${\psi o}_{1}  $, ${\psi o}_{2} $ and  ${\psi o}_{3}$ are the enabling mappings of  ${\varphi o}_{1}$, ${\varphi o}_{2}$ and ${\varphi o}_{3} $, respectively. This process establishes  $M $ as a typical finite automaton and an operational information system.

\subsection{Formal Modeling and Realization of Training Information}

The second stage in Fig. 2 is the realization of training information. According to the general formalization of information, its formal expression is straightforward:
\begin{equation}
It:S_{ot } (ot,Tt_h) \rightrightarrows S_{M_I} (M_I,t_1)
\end{equation}
\begin{equation}
S_{ot}(ot,{Tt}_{h}) = \left\{ \varphi t \middle| \varphi t\ is\ a\ WFF\ about\ ot\ and\ {Tt}_{h} \right\}
\end{equation}
\begin{equation}
S_{M_{I}}\left( M_{I},t_{1} \right) = \left\{ \psi t \middle| \psi t\ is\ a\ WFF\ about\ M_{i}\ and\ t_{1} \right\}
\end{equation}

Here, the ontology $ot $ refers to the object containing knowledge or samples;  ${Tt}_{h} $ is the time at which its state is formed; and $S_{ot}(ot,{Tt}_{h}) $ is a set of business-oriented semantic assertions or facts about the ontology. $M_{I} $ functions similarly to the hardware interface and API software of  $M $, carrying the training information at time $t_{1} $. The carrier state  $S_{M_{I}}\left( M_{I},t_{1} \right) $ is the physical realization of $S_{ot}(ot,{Tt}_{h}) $ via an enabling mapping, ensuring that the information can be machine-read and entered into $M $. In practice, much training information falls under the noisy information defined in Definition 12. The structure of noisy information will affect the effectiveness of model training.

\subsection{Formal Modeling and Implementation of Model Learning}

In the third stage of Fig. 2, $M $’s learning is precisely the fusion of the original information $Io $ with the training information $It $, optimizing it into updated information:
\begin{equation}
Iu:S_{M \cup M_{I} } (M \cup M_{I},t_{1}) \rightrightarrows S_{M} (M,t_2)
\end{equation}

Here, the ontology is the hybrid  $M \cup M_{I} $, encompassing both the model and its interface;  $t_{1} \cup t_{2} $  marks the time of the first learning event; and  $S_{M \cup M_{I}}(M \cup M_{I},t_{1} \cup t_{2}) $ formally represents the hybrid state during learning:
\begin{equation}
S_{M \cup M_{I}}(M \cup M_{I},t_{1} \cup t_{2}) = \left\{ {\varphi u}_{1},{\varphi u}_{2},{\varphi u}_{3} \right\}
\end{equation}
\begin{equation}
{\varphi u}_{1} = {Eq}^{2}(\Phi^{2}\left( M,t_{2} \right),f_{learn}^{2}(M,S_{M_{I}}\left( M_{I},t_{1} \right))) \land {State}^{3}\left( M,t_{2},\Phi^{2}(M,t_{2}) \right)
\end{equation}
indicates that the input state  $S_{M_{I}}\left( M_{I},t_{1} \right) $, which was learned by  $M $ at time  $t_{1} $, has been transformed into a new set of structures, parameters, and algorithms  $\Phi^{2}(M,t_{2}) $ at time  $t_{2} $, thus becoming the state of  $M $ at  $t_{2} $. As shown in Equations (4.3) and (4.4), $\varphi_{u_2}$ and $\varphi_{u_3}$ are logically identical to $\varphi_{o_2}$ and $\varphi_{o_3}$, respectively, differing only in the underlying ontological object and time. This indicates that the state retains the capacity for continued learning and processing even after the initial learning and optimization phases.

\subsection{User Input Information for the Model}

User input information in the fourth stage in Fig. 2 is formally given by:
\begin{equation}
Iq:S_{oq } (oq,Tq_h) \rightrightarrows S_{M_I} (M_I,t_2)
\end{equation}
\begin{equation}
S_{oq}(oq,{Tq}_{h}) = \left\{ \varphi q \middle| \varphi q\ is\ a\ WFF\ about\ oq\ and\ {Tq}_{h} \right\}
\end{equation}
\begin{equation}
S_{M_{I}}(M_{I},t_{2}) = \left\{ \psi q \middle| \psi q\ is\ a\ WFF\ about\ M_{I}\ and\ t_{2} \right\}
\end{equation}

Here, the ontology  $oq $   represents the user's intention or device, with  ${Tq}_{h} $ denoting the time at which the ontological state is formed. The state  $S_{oq}(oq,{Tq}_{h}) $ is the formal representation of the user's needs. The carrier state $S_{M_{I}}(M_{I},t_{2}) $ is the physical realization, via enabling mapping, of  $S_{oq}(oq,{Tq}_{h}) $ onto the model's interface device $M_{I} $. User input information is also typically affected by noise. Its noisy information structure will be directly related to the $M $’s output error.

\subsection{Information Processing, Output, and Online Learning of the Model}

In the fifth stage shown in Fig. 2, when the model $M $ processes input and produces output, the information it carries can be represented as:
\begin{equation}
Iv:S_{M \cup M_{I}\cup M_{O} } (M \cup M_{I}\cup M_{O},t_{2}\cup t_{3}) \rightrightarrows S_{M} (M,t_3)
\end{equation}
where:
\begin{equation}
S_{M \cup M_{I} \cup M_{O}}\left( M \cup M_{I} \cup M_{O},t_{2} \cup t_{3} \right) = \left\{ {\varphi v}_{1},{\varphi v}_{2} \right\}
\end{equation}

$M_O$ is $M$’s output device. The formulas $\varphi_{v_1}$ and $\varphi_{v_2}$ are logically identical to $\varphi_{u_2}$ and $\varphi_{u_3}$, respectively, differing only in their temporal references. This demonstrates that if the input information is processable and learnable by $M$, then $M$ can perform online learning concurrently with information processing and output operations.

The output information is given by:
\begin{equation}
Ir:S_{M \cup M_{I}\cup M_{O} } (M \cup M_{I}\cup M_{O},t_{2}) \rightrightarrows S_{M_{O}} (M_{O},t_3)
\end{equation}
where the ontological state is:
\begin{equation}
S_{M \cup M_{I}\cup M_{O}}\left( M \cup M_{I}\cup M_{O},t_{2} \right) = \left\{ {\varphi r}_{1} \right\}
\end{equation}
with
\begin{equation}
{\varphi r}_{1} = {Eq}^{2}(S_{M_{O}}\left( M_{O},t_{3} \right),f_{process}^{2}(M,S_{M_{I}}(M_{I},t_{2})))
\end{equation}

Finally, it should be emphasized that $M$ can repeat the processes described in Sections 4.2 to 4.5, iterating through cycles of training and information processing to meet user demands and continuously improve its performance.

\section{Several Important Definitions and Theorems}

The MLT-MF framework is applicable to a range of key issues in the field of artificial intelligence, offering many unique and practical theoretical principles and methodological guidelines.

\subsection{Theorem of Model Interpretability}

According to the view that interpretability should be a concept dependent on the domain of discourse and the corresponding logical system, and that interpretability should also be related to the specific inputs and outputs of the model \citep{glanois2024survey}, the following concepts are established based on the MLT-MF framework:

\textbf{Definition 13 (Model Interpretability)} Let the model  $M $, as a finite automaton, operate over an ordered set of times  $T = \left\{ t_{i} \middle| i = 1,\ldots,n \right\} $, at which it undergoes assignment, input, output, and state transitions. The state of  $M $ at time  $t_{i} $  is denoted as  $S_{M}\left( M,t_{i} \right) $, and the state of its input device  $M_{I} $  at time  $t_{i} $  is  $S_{M_{I}}\left( M_{I},t_{i} \right) $, while the state of the output device  $M_{O} $  at time  $t_{i + 1} $  is
 $S_{M_{O}}\left( M_{O},t_{i + 1} \right) $. If there exists a formal system  $\mathcal{L} $ along with three interpretations  $E_{M} $,  $E_{M_{I}} $, and  $E_{M_{O}} $  over the finite domains  $D_{M} $,  $D_{M_{I}} $, and  $D_{M_{O}} $  respectively, such that
\begin{equation}
E_{M_{O}} \subseteq Cn(E_{M} \cup E_{M_{I}})
\end{equation}
making  $S_{M}\left( M,t_{i} \right) $, $S_{M_{I}}\left( M_{I},t_{i} \right) $, and $S_{M_{O}}\left( M_{O},t_{i + 1} \right) $ the respective enabling mappings of  $E_{M} $, $E_{M_{I}} $, and $E_{M_{O}} $, then  $M $ is said to be interpretable in  $\mathcal{L} $ with respect to output $S_{M_{O}}\left( M_{O},t_{i + 1} \right) $ under input $S_{M_{I}}\left( M_{I},t_{i} \right) $. If for all $i = 1,\cdots,n - 1 $, the output $S_{M_{O}}\left( M_{O},t_{i + 1} \right) $ under input $S_{M_{I}}\left( M_{I},t_{i} \right) $ is interpretable in $\mathcal{L} $, then  $M $ is called interpretable in  $\mathcal{L} $.

Here,  $Cn(E_{M} \cup E_{M_{I}}) $ denotes the logical closure of $E_{M} \cup E_{M_{I}} $. Equation (5.1) states that in the formal system  $\mathcal{L} $, all formulas in $E_{M_{O}} $   are logical consequences of $E_{M} $  and  $E_{M_{I}} $.Moreover, since  $S_{M}\left( M,t_{i} \right) $, $S_{M_{I}}\left( M_{I},t_{i} \right) $, and $S_{M_{O}}\left( M_{O},t_{i + 1} \right) $ are respectively enabled by $E_{M} $,  $E_{M_{I}} $, and $E_{M_{O}} $, the formal system  $\mathcal{L} $ and its three interpretations fully capture the logical and causal relationship whereby  $M $ produces output $S_{M_{O}}\left( M_{O},t_{i + 1} \right) $ under input $S_{M_{I}}\left( M_{I},t_{i} \right) $. This, in essence, offers a concise understanding of model interpretability.

\textbf{Theorem 1 (Model Interpretability and Information Existence)}
Let the model  $M $ be viewed as a finite automaton operating over an ordered set of times $T = \left\{ t_{i} \middle| i = 1,\ldots,n \right\} $, during which it undergoes assignment, input, output, and state transitions.

{\sloppy\emergencystretch=1em
(1) For any  \(i=1,\dots,n-1\), a necessary and sufficient condition for the output  $S_{M_{O}}\left( M_{O},t_{i + 1} \right) $ of model  $M $ to be interpretable in the formal system  $\mathcal{L} $ under the input  $S_{M_{I}}\left( M_{I},t_{i} \right) $ is that there exist  $Iu(t_i): S_{ou}(ou(t_i), Tu_{h}(t_i)) \rightrightarrows S_M(M, t_i)$  and  $Iq(t_i): S_{oq}(oq(t_i), Tq_{h}(t_i)) \rightrightarrows S_{M_I}(M_I, t_i)$  representing the information carried by  $M $ and its input device  $M_{I} $  at time  $t_{i} $, respectively, and  $Ir(t_{i+1}): S_{or}(or(t_{i+1}), Tr_{h}(t_{i+1})) \rightrightarrows S_{M_O}(M_O, t_{i+1})$  representing the information carried by the output device  $M_{O} $ at time  $t_{i + 1} $. The ontological states of these three pieces of information must each correspond to an interpretation in $\mathcal{L} $, and $S_{or}(or(t_{i + 1}),{Tr}_{h}(t_{i + 1})) \subseteq Cn(S_{ou}(ou(t_{i}),{Tu}_{h}(t_{i})) \cup S_{oq}(oq(t_{i}),{Tq}_{h}(t_{i}))) $ must hold.
\par}

(2) A necessary and sufficient condition for  $M $ to be interpretable in  $\mathcal{L} $ is that, for all  $i = 1,\cdots,n - 1 $, there exist information objects  $Iu(t_{i}) $,  $Iq(t_{i}) $, and  $Ir(t_{i + 1}) $ whose ontological states are all interpretations in  $\mathcal{L} $, and such that the ontological state of  $Ir(t_{i + 1}) $ is contained in the logical closure of the union of the ontological states of  $Iu(t_{i}) $ and $Iq(t_{i}) $.

\textbf{Proof:} For any \(i=1,\dots,n-1\), suppose that the output $S_{M_{O}}\left( M_{O},t_{i + 1} \right) $ of model  $M $ under the input  $S_{M_{I}}\left( M_{I},t_{i} \right) $ is interpretable in the formal system  $\mathcal{L} $. According to Definition 13, there exist three interpretations  $E_{M} $,  $E_{M_{I}} $, and $E_{M_{O}} $   in the formal system  $\mathcal{L} $, respectively over the finite domains  $D_{M} $,  $D_{M_{I}} $, and  $D_{M_{O}} $, such that $E_{M_{O}} \subseteq Cn(E_{M} \cup E_{M_{I}}) $, and $S_{M}\left( M,t_{i} \right) $, $S_{M_{I}}\left( M_{I},t_{i} \right) $, and $S_{M_{O}}\left( M_{O},t_{i + 1} \right) $ are each the enabling mappings of  $E_{M} $,  $E_{M_{I}} $, and $E_{M_{O}} $, respectively.

{\sloppy\emergencystretch=1em
Let \(S_{ou}(ou(t_{i}),Tu_{h}(t_{i}))=E_{M}\), \(S_{oq}(oq(t_{i}),Tq_{h}(t_{i}))=E_{M_{I}}\) and \(S_{or}(or(t_{i+1}),Tr_{h}(t_{i+1}))=E_{M_{O}}\). Then, according to Definition 13, there exist information objects $Iu(t_{i}): S_{ou}(ou(t_{i}), Tu_{h}(t_{i})) \rightrightarrows S_{M}(M, t_{i})$, $Iq(t_{i}): S_{oq}(oq(t_{i}), Tq_{h}(t_{i})) \rightrightarrows S_{M_I}(M_I, t_{i})$ and $Ir(t_{i+1}): S_{or}(or(t_{i+1}), Tr_{h}(t_{i+1})) \rightrightarrows S_{M_O}(M_O, t_{i+1})$ such that \(S_{or}(or(t_{i+1}),Tr_{h}(t_{i+1}))\subseteq Cn(S_{ou}(ou(t_{i}),Tu_{h}(t_{i}))\cup S_{oq}(oq(t_{i}),Tq_{h}(t_{i})))\). Thus, the necessity of the first part of the theorem is established.
\par}

{\sloppy\emergencystretch=1em
On the other hand, for any  $1 \leq i \leq n - 1 $, suppose there exist information objects  $Iu(t_{i}): S_{ou}(ou(t_{i}), Tu_{h}(t_{i})) \rightrightarrows S_{M}(M, t_{i})$,  $Iq(t_{i}): S_{oq}(oq(t_{i}), Tq_{h}(t_{i})) \rightrightarrows S_{M_I}(M_I, t_{i})$ and $Ir(t_{i+1}): S_{or}(or(t_{i+1}), Tr_{h}(t_{i+1})) \rightrightarrows S_{M_O}(M_O, t_{i+1})$,
such that in the formal system  $\mathcal{L} $, 
$S_{or}(or(t_{i + 1}),{Tr}_{h}(t_{i + 1}))\subseteq Cn(S_{ou}(ou(t_{i}),{Tu}_{h}(t_{i})) \cup S_{oq}(oq(t_{i}),{Tq}_{h}(t_{i})))
$. Then, according to Lemma A.1, $E_{M} = S_{ou}(ou(t_{i}),{Tu}_{h}(t_{i})) $, $E_{M_{I}} = S_{oq}(oq(t_{i}),{Tq}_{h}(t_{i})) $, and $E_{M_{O}} = S_{or}\left( or\left( t_{i + 1} \right),{Tr}_{h}\left( t_{i + 1} \right) \right) $ are interpretations of  $\mathcal{L} $ over the three finite domains $ou(t_{i}) \times {Tu}_{h}(t_{i}) $, $oq(t_{i}) \times {Tq}_{h}(t_{i}) $, and $r(t_{i + 1}) \times {Tr}_{h}(t_{i + 1}) $, respectively, and $E_{M_{O}} \subseteq Cn(E_{M} \cup E_{M_{I}}) $. At the same time, $S_{M}\left( M,t_{i} \right) $, $S_{M_{I}}\left( M_{I},t_{i} \right) $, and $S_{M_{O}}\left( M_{O},t_{i + 1} \right) $ are the enabling mappings of $E_{M} $,  $E_{M_{I}} $, and $E_{M_{O}} $, respectively. By Definition 13, the output  $S_{M_{O}}\left( M_{O},t_{i + 1} \right) $ of model  $M$ under the input  $S_{M_{I}}\left( M_{I},t_{i} \right) $ is interpretable
in the formal system  $\mathcal{L} $. Therefore, the sufficiency of the first part of the theorem is established.
\par}

For all  $i = 1,\cdots,n - 1 $, by combining Definition 13 with the first part of the theorem, the second part of the theorem follows directly. This completes the proof.

Definition 13 provides a rigorous mathematical definition of model interpretability. Furthermore, Theorem 1 demonstrates that, throughout the processes of model construction and application, as long as the ontological states of the model and its input and output information can be established within the same formal system, it is possible to reveal the causal relationships between model inputs and outputs within this system, thereby achieving model interpretability. It should also be noted that formal systems vary greatly across different academic disciplines, and the myriad entities in the world can be expressed through information in the forms of formulas, code, text, graphics, or even video. Moreover, the same state can be represented in different information modalities according to varying user requirements. Therefore, Definition 13 and Theorem 1 theoretically establish a foundation for the flexible use of multiple information modalities and for providing model interpretability tailored to user needs.

\subsection{Theorem of Ethical Safety Assurance}

To apply the MLT-MF framework, it is first necessary to identify the corresponding well-formed ethical constraint formulas  $Ec$ in accordance with fundamental ethical standards such as fairness and privacy protection, thereby establishing a formalized expression of ethical security.

\textbf{Definition 14 (Ethical Safety)} Let  $S(X,T) $ be the interpretation of a set of WFFs in a formal system  $\mathcal{L} $ over the domain  $X \times T $. Let  $Ec(\overrightarrow{v}) $ denote an ethical constraint expressed as a WFF in  $\mathcal{L} $, where  $\overrightarrow{v} = (v_{1},\ldots,v_{n}) $ are its free variables. For each substitutable  $(x,t) \in X \times T $, define
\begin{equation}
Ec|_{(x,t)} = Ec(\overrightarrow{v} \mapsto \overrightarrow{\sigma}(x,t))
\end{equation}
where  $\overrightarrow{\sigma}:X \times T \rightarrow D^{n} $ is a value vector function such that  $\sigma_{i}(x,t) \in D $ is the concrete value assigned to  $v_{i} $ for  $1 \leq i \leq n $.

If for all  $\varphi \in Cn(S(X,T)) $, and all  $(x,t) \in X \times T $,
\begin{equation}
\varphi \rightarrow \sim Ec|_{(x,t)}
\end{equation}
does not hold, then  $S(X,T) $ is said to be ethically safe with respect to  $Ec $. Furthermore, if a model M is capable of producing output  $I $, and both the ontological and carrier states of  $I $ are ethically safe with respect to  $Ec $, then M is said to be ethically safe with respect to  $Ec $.

In the definition of ethical safety, we require not only that  $S(X,T) $ contains no formulas that entail  $\sim Ec|_{(x,t)} $, but also that its logical closure  $Cn(S(X,T)) $ contains no such formulas. This is because if  $Cn(S(X,T)) $ were to include formulas entailing  $\sim Ec|_{(x,t)} $, then the conjunction of multiple formulas in  $S(X,T) $ could still violate the ethical constraint  $Ec $. This would indicate that  $S(X,T) $ implicitly allows situations that contravene  $Ec $, which must be strictly avoided in a rigorous sense. Moreover, it is generally not difficult for a model M to achieve ethical safety with respect to  $Ec $ in a simple manner; the key challenge lies in ensuring the largest possible set of valid outputs while maintaining ethical safety. To this end, we propose:

\textbf{Theorem 2 (Ethical Safety Assurance)} Suppose model M is capable of processing  $Iq $ with  $Iu $ and outputting  $Ir $. Let  $S_{or}\left( or,{Tr}_{h} \right) $ denote the ontological state of  $Ir $. For any ethical constraint  $Ec $ represented by a well-formed formula in  $\mathcal{L} $, there exists a constructive algorithm for obtaining the largest output state subset  $S_{or}^{Ec}\left( or,{Tr}_{h} \right) \subseteq S_{or}\left( or,{Tr}_{h} \right) $ that is ethically safe with respect to  $Ec $. By introducing an ethical safeguard formula for  $Ec $ into  $Iu $, it is thereby possible to ensure that model M attains the maximal output state that is ethically safe with respect to  $Ec $.

\textbf{Proof:} Since model M is capable of processing  $Iq $ with $Iu $ and outputting  $Ir $, for the ontological states  $S_{ou}\left( ou,{Tu}_{h} \right) $, $S_{oq}\left( oq,{Tq}_{h} \right) $ and  $S_{or}\left( or,{Tr}_{h} \right) $ of  $Iu $, $\ Iq $ and  $Ir $ respectively, the MLT-MF framework provides:
\begin{equation}
\begin{aligned}
\varphi u = &\bigl((\forall\,\Phi u)(\forall\,\Phi q)\bigl(
  {IsSubsetOf}^{2}(\Phi u,S_{ou}(ou,{Tu}_{h})) \\
&\quad \land {IsSubsetOf}^{2}(\Phi q,S_{oq}(oq,{Tq}_{h})) \\
&\quad \land {Processable}^{2}(\Phi u,\Phi q)
\bigr)\bigr) \\
&\rightarrow \bigl(\exists\,\varphi_r\bigl(
  {IsFormulaOf}^{2}(\varphi r,S_{or}(or,{Tr}_{h})) \\
&\quad \land {Eq}^{2}(\varphi r,f_{\mathrm{process}}^{2}(\Phi u,\Phi q))
\bigr)\bigr).
\end{aligned}
\end{equation}

Thus, the set
\begin{equation}
\begin{split}
S_{or}\left( or,{Tr}_{h} \right) = \left\{ \varphi r \middle| \right. & \left. \varphi r = f_{\mathrm{process}}^{2}(\Phi u,\Phi q), \right. \\
&\left. \Phi u \in Sou\left( ou,{Tu}_{h} \right), \right. \\
&\left. \Phi q \in S_{oq}\left( oq,{Tq}_{h} \right) \right\}
\end{split}
\end{equation}
represents the theoretical output states of model $M$.

Since  $S_{or}\left( or,{Tr}_{h} \right) $ is a finite set, we construct a finite hypergraph
\begin{equation}
H_{or} = \left( S_{or}\left( or,{Tr}_{h} \right),\mathcal{E} \right)
\end{equation}
where the vertex set is precisely  $S_{or}\left( or,{Tr}_{h} \right) $, and the hyperedge set  $\mathcal{E} $ covers all WFFs that directly or implicitly violate the ethical constraint  $Ec $:

(1) For any  $\varphi r \in S_{or}\left( or,{Tr}_{h} \right) $ and $(x,t) \in or \times {Tr}_{h} $, if $\varphi r \rightarrow \sim Ec|_{(x,t)} $, then add the
hyperedge  $\left\{ \varphi r \right\} $ to  $\mathcal{E} $;

(2) If there exists a minimal subset
 $W \subseteq S_{or}\left( or,{Tr}_{h} \right) $ such that the conjunction of all formulas in  $W $ entails  $\sim Ec|_{(x,t)} $, then add the hyperedge  $W $ to  $\mathcal{E} $.

Let  $G $ be an independent set in  $H_{or} $, i.e., a subset of vertices containing no hyperedge  $e \in \mathcal{E} $. If there exists
 $\varphi \in Cn(G) $ such that
\begin{equation}
\varphi \rightarrow \sim Ec|_{(x,t)}
\end{equation}
then there must exist a finite subset $\left\{ {\varphi r}_{1},\cdots,{\varphi r}_{k} \right\} \subseteq G $ such that ${\varphi r}_{1} \land \cdots \land {\varphi r}_{k} = \varphi $, and hence ${\varphi r}_{1} \land \cdots \land {\varphi r}_{k} \rightarrow \sim Ec|_{(x,t)} $. Let $W = \left\{ {\varphi r}_{1},\cdots,{\varphi r}_{k} \right\} $, so the conjunction of  $W $ entails  $\sim Ec|_{(x,t)} $.

According to the hyperedge construction rules, if  $|W| = 1 $, then $W $ itself directly violates  $Ec $, and there must be a hyperedge $\left\{ {\varphi r}_{1} \right\} $. But since $\left\{ {\varphi r}_{1} \right\} \in G $, this contradicts the assumption that  $G $ is an independent set. If  $|W| \geq 2 $, then by the compactness theorem, the logical entailment can be captured by a finite minimal subset, so there exists a minimal subset $W_{\min} \subseteq W $ whose conjunction entails $\sim Ec|_{(x,t)} $. Therefore, there exists a hyperedge $W_{\min} \subseteq W \subseteq G $ in  $H_{or} $, again contradicting the independence of  $G $. Thus, (5.7) cannot hold, and the independent set  $G $ is ethically safe with respect to  $Ec $.

By the maximum independent set theorem for finite hypergraphs\citep{Berge1973Graphs}, $H_{or} $  necessarily contains an independent set $G_{\max} $  of maximal cardinality. Suppose there exists another ethically safe subset  $K $ with
\begin{equation}
|K| > \left| G_{\max} \right|
\end{equation}
then  $K $ cannot contain any hyperedge  $e \in \mathcal{E} $. Otherwise, if $|e| = 1 $, then  $K $ directly violates  $Ec $; if  $|e| \geq 2 $, then the formulas in  $K $ violate  $Ec $. Both cases contradict the ethical safety assumption for  $K $. Therefore,  $K $ must also be an independent set in  $H_{or} $, and (5.8) cannot hold. Hence,  $G_{\max} $  is the largest ethically safe subset of  $S_{or}\left( or,{Tr}_{h} \right) $ with respect to  $Ec $.

Let  $S_{or}^{Ec}\left( or,{Tr}_{h} \right) = G_{\max} $ and $S_{or}^{\sim Ec}\left( or,{Tr}_{h} \right) = S_{or}\left( or,{Tr}_{h} \right){\backslash S}_{or}^{Ec}\left( or,{Tr}_{h} \right) $. In addition, introduce a new constant in  $\mathcal{L} $: NEc, representing a non-Ec ethical safety prompt, and a new predicate:  ${IsSafeback}^{1}(NEc) $, indicating that M outputs the non-Ec ethical safety prompt NEc. We may safely assume that  $\mathcal{L} $ originally contains neither.

In this way, add to  $S_{ou}\left( ou,{Tu}_{h} \right) $ an ethical safeguard formula:
\begin{equation}
\begin{split}
\varphi u = & (\forall\varphi r)\left( \mathit{IsFormulaOf}^{2}\left( \varphi r, S_{or}\left( or, Tr_{h} \right) \right) \right) \rightarrow \\
& \exists\varphi s\left( \mathit{IsFormulaOf}^{2}\left( \varphi s, S_{os}\left( os, Ts_{h} \right) \right) \land \right. \\
& \left( \mathit{IsFormulaOf}^{2}\left( \varphi r, S_{or}^{Ec}\left( or, Tr_{h} \right) \right) \rightarrow \mathit{Eq}^{2}(\varphi s, \varphi r) \right) \land \\
& \left. \left( \mathit{IsFormulaOf}^{2}\left( \varphi r, S_{or}^{\sim Ec}\left( or, Tr_{h} \right) \right) \rightarrow \mathit{Eq}^{2}\left( \varphi s, \mathit{IsSafeback}^{1}(NEc) \right) \right) \right)
\end{split}
\end{equation}
Since the newly added  $\varphi u $ pertains to $S_{os}\left( os,{Ts}_{h} \right) $ and is entirely independent of the existing formulas in  $S_{ou}\left( ou,{Tu}_{h} \right) $, it does not affect the satisfaction of Postulate A.1 by $S_{ou}\left( ou,{Tu}_{h} \right) $. Moreover,
\begin{equation}
S_{os}\left( os,{Ts}_{h} \right) = S_{or}^{Ec}\left( or,{Tr}_{h} \right) \cup \left\{ {IsSafeback}^{1}(NEc) \right\}
\end{equation}

{\sloppy\emergencystretch=1em
Because NEc and  ${IsSafeback}^{1}(NEc) $ were not previously present in $\mathcal{L} $, and  $S_{or}^{Ec}\left( or,{Tr}_{h} \right) $ and $\left\{ {IsSafeback}^{1}(NEc) \right\} $ are logically independent and both logically consistent,  $S_{os}\left( os,{Ts}_{h} \right) $ is also logically consistent and satisfies all other conditions of Postulate A.1. Therefore, it is the maximal output state that is ethically safe with respect to $Ec $.
\par}

By taking  $S_{os}\left( os,{Ts}_{h} \right) $ as the ontological state and mapping it via noise-free information  $Is $ to the carrier state  $S_{M}(M,{Ts}_{m}) $,  $S_{M}(M,{Ts}_{m}) $ is necessarily also ethically safe with respect to  $Ec $. This satisfies the condition in Definition 14 for model $M$ to be ethically safe with respect to  $Ec $. The theorem is thus proved.

Theorem 2 not only establishes that it is possible to construct an ethical safety assurance mechanism for models, but also provides a constructive algorithm—based on hypergraphs—for obtaining the SafeMax. However, computing the maximum independent set in a hypergraph is generally an NP-hard problem. In practical applications, the algorithm can be simplified by leveraging two functions: one for ranking the conflict intensity among formulas in $S_{or}\left( or,{Tr}_{h} \right) $, and another for ethical safety checking, thus yielding an approximately SafeMax. This is a practical issue that warrants further investigation and will not be elaborated upon here. Meanwhile, the inheritance of ethical safety in the learning process can be achieved by incorporating dedicated ethical constraint regularization terms into the model algorithm. This involves more specific model structures and algorithmic design, which are beyond the scope of this paper.

\subsection{Theorem on TVD Upper Bounds}

Applying the MLT-MF framework allows us to estimate upper bounds on the generalization error measured by TVD between model distributions, facilitating the analysis and prediction of model generalization performance across a variety of tasks.

\textbf{Theorem 3 (TVD Upper Bound for Conservative Prediction Strategies in Noise-Free Inputs and Uncovered Training Regions)} Let the information describing the model $M$ after learning and optimization be $Iu: S_{ou}(ou, Tu_h) \rightrightarrows S_M(M, Tu_m)$, and the input information be $Iq: S_{oq}(oq, Tq_h) \rightrightarrows S_{cq}(cq, Tq_m)$. Assume both $S_{ou}$ and $S_{oq}$ are distributed according to probability distributions over the same measurable space. Let $|S_{oq}|$ and $|S_{ou} \cap S_{oq}|$ denote the cardinalities of the finite sets $S_{oq}$ and $S_{ou} \cap S_{oq}$, respectively. The probability distribution of the model is $p_M(x)$, where $p_M(x) > 0$ for $x \in S_{ou}$; the distribution over $S_{oq}$ is $p_q(x)$, with $p_q(x) > 0$ for $x \in S_{oq}$. The total probability mass of $p_q(x)$ over $S_{ou} \cap S_{oq}$ is denoted $P_{Mq}$. When the model $M$ is sufficiently trained on $S_{ou}$, and adopts a conservative uniform prediction for unknown states, the total variation distance (TVD) between $p_M(x)$ and $p_q(x)$ satisfies:
\begin{align}
\mathrm{TVD}(p_q, p_M) = 
\begin{cases}
\approx 0, & \text{if } S_{oq} \subset S_{ou} \\[1em]
\leq \sqrt{\dfrac{1}{2}D_{\text{KL}}}, & \text{if } S_{oq} \neq S_{ou} \cap S_{oq}
\end{cases}
\end{align}
\text{where }$D_{\text{KL}} = (1\!-\!P_{\!Mq})\log\!\left(\!\dfrac{|S_{oq}|\!-\!|S_{ou} \cap S_{oq}|}{1\!-\!P_{\!Mq}}\!\right)\! -\! H(p_q)$.

For the second case, it is assumed that $S_{oq}\backslash S_{ou} \neq \varnothing $,  $1 - P_{Mq} > 0 $, and $\left| S_{oq} \right| - \left| S_{ou} \cap S_{oq} \right| > 0 $. Here, $H(p_{q}) = - \sum_{x \in S_{oq}\backslash S_{ou}}^{}{p_{q}(x)\log p_{q}(x)} $ is the entropy of the data in the non-overlapping region.

\textbf{Proof:} Since  $M $ is sufficiently trained on $S_{oq} $  within the overlapping region $S_{ou} \cap S_{oq} $, and adopts a conservative uniform distribution to predict unknown states,  $p_{M}(x) $ closely approximates $p_{q}(x) $ in the overlapping region  $S_{ou} \cap S_{oq} $, and follows a uniform distribution over the non-overlapping region $S_{oq}\backslash S_{ou} $. That is,
\begin{equation}
p_{M}(x) = \begin{cases}
\approx p_{q}(x), & \text{if } x \in S_{ou} \cap S_{oq} \\[0.3em]
\frac{1 - P_{Mq}}{|S_{oq}| - |S_{ou} \cap S_{oq}|}, & \text{if } x \in S_{oq}\backslash S_{ou}
\end{cases}
\end{equation}

Using KL-divergence to measure the difference between the true distribution and the hypothesized model distribution, we have \citep{goodfellow2016deep}:
\begin{align}
D_{KL}(p_{q}(x) \parallel p_{M}(x)) &= \sum_{x \in S_{oq}} p_{q}(x)\log\frac{p_{q}(x)}{p_{M}(x)} \nonumber\\
&= \sum_{x \in S_{ou} \cap S_{oq}} p_{q}(x)\log\frac{p_{q}(x)}{p_{M}(x)} + \sum_{x \in S_{oq}\backslash S_{ou}} p_{q}(x)\log\frac{p_{q}(x)}{p_{M}(x)}
\end{align}

According to (5.12), in the overlapping region,
\begin{align}
\sum_{x \in S_{ou} \cap S_{oq}} p_{q}(x)\log\frac{p_{q}(x)}{p_{M}(x)} &\approx \sum_{x \in S_{ou} \cap S_{oq}} p_{q}(x)\log\frac{p_{q}(x)}{p_{q}(x)} \nonumber\\
&= 0
\end{align}

This establishes the first case in (5.11).

In the non-overlapping region, the true distribution  $p_{q}(x) $ can be
arbitrary. Similarly, by (5.12), we have
\begin{align}
\sum_{x \in S_{oq}\backslash S_{ou}} p_{q}(x)\log\frac{p_{q}(x)}{p_{M}(x)} &= \sum_{x \in S_{oq}\backslash S_{ou}} p_{q}(x)\log p_{q}(x) - \sum_{x \in S_{oq}\backslash S_{ou}} p_{q}(x)\log p_{M}(x) \nonumber\\
&= -H(p_{q}) - \sum_{x \in S_{oq}\backslash S_{ou}} p_{q}(x)\log\frac{(1 - P_{Mq})}{|S_{oq}| - |S_{ou} \cap S_{oq}|} \nonumber\\
&= -H(p_{q}) - \log\frac{(1 - P_{Mq})}{|S_{oq}| - |S_{ou} \cap S_{oq}|}\sum_{x \in S_{oq}\backslash S_{ou}} p_{q}(x) \nonumber\\
&= (1 - P_{Mq})\log\frac{|S_{oq}| - |S_{ou} \cap S_{oq}|}{1 - P_{Mq}} - H(p_{q})
\end{align}

Since both  $p_{M}(x) $ and  $p_{q}(x) $ are probability distributions
defined on the same measurable space, we can use KL-divergence to
estimate an upper bound on their TVD. By Pinsker's
inequality \citep{pinsker1964information}, we have:
\begin{align}
TVD(p_{q},p_{M}) &\leq \sqrt{\frac{1}{2}D_{KL}(p_{q}(x) \parallel p_{M}(x))} \nonumber\\
&= \sqrt{\frac{1}{2}\left(\sum_{x \in S_{ou} \cap S_{oq}} p_{q}(x)\log\frac{p_{q}(x)}{p_{M}(x)} + \sum_{x \in S_{oq}\backslash S_{ou}} p_{q}(x)\log\frac{p_{q}(x)}{p_{M}(x)}\right)} \nonumber\\
&\approx \sqrt{\frac{1}{2}\left((1 - P_{Mq})\log\frac{|S_{oq}| - |S_{ou} \cap S_{oq}|}{1 - P_{Mq}} - H(p_{q})\right)}
\end{align}

This establishes the second case in (5.11).

Since we have already assumed that the carrier states $S_M (M, Tu_m)$ and $S_{cq} (cq, Tq_m)$ of $Iu: S_{ou} (ou, Tu_h) \rightrightarrows S_M (M, Tu_m)$ and $Iq: S_{oq} (oq, Tq_h) \rightrightarrows S_{cq} (cq, Tq_m)$ are both represent the physical realizations of $S_{ou}$ and $S_{oq}$ in model $M$ and its interface, respectively. It follows that equation (5.11) characterizes the upper bound of the model's TVD. This completes the proof.

Distinct from traditional generalization theories such as VC dimension and Rademacher complexity, Theorem 3, based on the MLT-MF framework, utilizes $P_{Mq}$ and $|S_{oq}| - |S_{ou} \cap S_{oq}|$---which reflect the overlap between test data and the model's original information---as well as the entropy $H(p_q)$ of test data in the non-overlapping region, as evaluation metrics to provide an explicit quantitative estimation of the TVD. Since TVD reflects the degree of distributional alignment of the model, independent of specific models and tasks, it is generally applicable for analyzing training performance. Moreover, for classification tasks with 0-1 loss, the upper bound of the generalization error is equivalent to the TVD \citep{rodriguez2021tighter}, while for regression tasks, the upper bound of the generalization error is given by the product of the TVD and the Lipschitz loss \citep{devroye2013probabilistic}. Therefore, TVD serves as an indicator for estimating a model's generalization error. In this way, Theorem 3 offers a novel perspective and practical guidance for model design and training data acquisition.

In general, input information is inevitably affected by noise, and arbitrary predictive strategies can be adopted in regions not covered by training data. According to Definition 12 on structural decomposition of noisy information, we can thus obtain more general results regarding the model's TVD.

\textbf{Theorem 4 (TVD Upper Bound for Arbitrary Prediction Strategies with Noisy Inputs and Uncovered Training Regions)} Let the base space $X $ be a finite set. The training information domain of the model  $M $ is  $S_{ou} \subseteq X $. The true problem domain is $S_{oq} \subseteq X $, and the true distribution  $p_{q} $ satisfies $\sum_{x \in S_{oq}}^{}{p_{q}(x)} = 1 $. The noisy information ontological state  $\widetilde{S_{oq}} = S_{oq}\backslash S_{oq}^{-} \cup S_{oq}^{+} $, as defined in Definition 12, replaces  $S_{oq} $ as the actual input to the model  $M $. The noise loss probability mass is $P_{l} = \sum_{x \in S_{oq}^{-}}^{}{p_{q}(x)} \in \lbrack 0,1\rbrack $, and the noise addition probability mass is $P_{a} = \sum_{x \in S_{oq}^{+}}^{}{p^{+}(x)} \in \lbrack 0,1\rbrack $, where  $p^{+} $ is a non-negative kernel on the noise addition region, satisfying  $1 - P_{l} + P_{a} > 0 $. The distribution of the noisy input  $\widetilde{S_{oq}} $ is given by $\widetilde{p_{q}}(x) = (p_{q}(x) \cdot 1\left\{ x \notin S_{oq}^{-} \right\} + p^{+}(x) \cdot 1\left\{ x \in S_{oq}^{+} \right\})/(1 - P_{l} + P_{a})$. Let  $V = \widetilde{S_{oq}}\backslash S_{ou} $ denote the region not covered by training information, with uncovered probability mass $P_{uncov} = \sum_{x \in V}^{}{\widetilde{p_{q}}(x)} \in \lbrack 0,1\rbrack $. The conditional distribution on  $V $ is $r(x) = \widetilde{p_{q}}(x)/P_{uncov} $  (and when $P_{uncov} = 0 $, all terms involving  $V $ below are taken as 0). The predictive strategy  $p_{0} $  on the uncovered region  $V $ is an arbitrary distribution, satisfying  $\sum_{x \in V}^{}{p_{0}(x) = 1,}\,p_{0}(x) \geq 0 $. The model distribution  $p_{M} $  is determined by training on the covered region  $U = \widetilde{S_{oq}} \cap S_{ou} $, and on  $V $,  $p_{M}(x) = P_{uncov}p_{0}(x) $. Assume that the model\textquotesingle s fitting error on the covered region is  $E_{fit} = D_{KL}\left( \widetilde{p_{q}}|_{U} \parallel p_{M}|_{U} \right) $, where  $\widetilde{p_{q}}|_{U} $  and
 $p_{M}|_{U} $  denote the distributions restricted to  $U $ and normalized (under the ideal conditions assumed in Theorem 3, $E_{fit} $  tends to zero).

Thus, the upper bound of the TVD for model  $M $ with respect to noisy input information is given by
\begin{align}
TVD\left( \widetilde{p_{q}},p_{M} \right) \leq \sqrt{\frac{\left\lbrack E_{fit} + P_{uncov}D_{KL}\left( r \parallel p_{0} \right) \right\rbrack}{2}} 
\end{align}

The upper bound of the TVD for model  $M $ with respect to the true information is
\begin{align}
TVD\left( p_{q},p_{M} \right) \leq \left\{ \begin{array}{r}
P_{l} + \sqrt{\frac{\left\lbrack E_{fit} + P_{uncov}D_{KL}\left( r \parallel p_{0} \right) \right\rbrack}{2}},\ \ \ \ \ \ \ \ \ \ \ if\ P_{l} \geq P_{a},\  \\
\frac{P_{a}}{1 - P_{l} + P_{a}} + \sqrt{\frac{\left\lbrack E_{fit} + P_{uncov}D_{KL}\left( r \parallel p_{0} \right) \right\rbrack}{2}},\ \ \ \ \ \ if\ P_{l} < P_{a}.
\end{array} \right.\ 
\end{align}

\textbf{Proof:} Based on the conditions provided in the theorem, we have:

 $D_{KL}\left( \widetilde{p_{q}} \parallel p_{M} \right) = \sum_{x \in \widetilde{S_{oq}}}^{}{\widetilde{p_{q}}(x)\log\frac{\widetilde{p_{q}}(x)}{p_{M}(x)}} $

 $= \sum_{x \in U}^{}{\widetilde{p_{q}}(x)\log\frac{\widetilde{p_{q}}(x)}{p_{M}(x)}} + \sum_{x \in V}^{}{\widetilde{p_{q}}(x)\log\frac{\widetilde{p_{q}}(x)}{p_{M}(x)}} $

 $= D_{KL}\left( \widetilde{p_{q}}|_{U} \parallel p_{M}|_{U} \right) + \sum_{x \in V}^{}{P_{uncov}r(x)\log\frac{P_{uncov}r(x)}{P_{uncov}p_{0}(x)}} $
\begin{equation}
= E_{fit} + P_{uncov}D_{KL}\left( r \parallel p_{0} \right)
\end{equation}

Here,  $E_{fit} $  reflects the degree of fit of  $M $ on the region covered by the training information, and its probabilistic upper bound can be characterized by measures such as VC dimension, Rademacher complexity, or training stability. Moreover, by applying Pinsker's inequality, the first conclusion of the theorem can be obtained:
\begin{equation}
TVD\left( \widetilde{p_{q}},p_{M} \right) \leq \sqrt{\frac{D_{KL}\left( \widetilde{p_{q}} \parallel p_{M} \right)}{2}} = \sqrt{\frac{\left\lbrack E_{fit} + P_{uncov}D_{KL}\left( r \parallel p_{0} \right) \right\rbrack}{2}}
\end{equation}

Since the distribution of the noisy input $\widetilde{S_{oq}} $  is
\begin{equation}
\widetilde{p_{q}}(x) = \frac{p_{q}(x) \cdot 1\left\{ x \notin S_{oq}^{-} \right\} + p^{+}(x) \cdot 1\left\{ x \in S_{oq}^{+} \right\}}{1 - P_{l} + P_{a}}
\end{equation}

It follows that
\begin{align}
\text{TVD}(p_q, \widetilde{p_q}) 
&= \frac{1}{2}\sum_{x \in X} \left| p_q(x) - \widetilde{p_q}(x) \right| \nonumber \\
&= \frac{1}{2} \bigg( \sum_{x \in S_{oq}^{-}} \left| p_q(x) - \widetilde{p_q}(x) \right| 
+ \sum_{x \in S_{oq}^{+}} \left| p_q(x) - \widetilde{p_q}(x) \right| \nonumber \\
&\quad + \sum_{x \in X \setminus (S_{oq}^{-} \cup S_{oq}^{+})} \left| p_q(x) - \widetilde{p_q}(x) \right| \bigg) \nonumber \\
&= \frac{1}{2} \bigg( \sum_{x \in S_{oq}^{-}} \left| p_q(x) \right| 
+ \sum_{x \in S_{oq}^{+}} \left| \frac{p^{+}(x)}{1 - P_l + P_a} \right| \nonumber \\
&\quad + \sum_{x \in X \setminus (S_{oq}^{-} \cup S_{oq}^{+})} \left| p_q(x) - \frac{p_q(x)}{1 - P_l + P_a} \right| \bigg) \nonumber \\
&= \frac{1}{2} \left( P_l + \frac{P_a}{1 - P_l + P_a} 
+ \frac{|P_l - P_a| \sum_{x \in X \setminus (S_{oq}^{-} \cup S_{oq}^{+})} p_q(x)}{1 - P_l + P_a} \right) \nonumber \\
&= \frac{1}{2} \left( P_l + \frac{P_a}{1 - P_l + P_a} 
+ \frac{|P_l - P_a| (1 - P_l)}{1 - P_l + P_a} \right) \nonumber \\
&= \begin{cases}
P_l, & \text{if } P_l \geq P_a, \\[2pt]
\dfrac{P_a}{1 - P_l + P_a}, & \text{if } P_l < P_a.
\end{cases}
\end{align}

By applying the triangle inequality, we have
\begin{equation}
\begin{aligned}
\text{TVD}(p_q, p_M) &\leq \text{TVD}(p_q, \widetilde{p_q}) + \text{TVD}(\widetilde{p_q}, p_M) \\
&\leq \begin{cases}
P_l + \sqrt{\frac{E_{fit} + P_{uncov} D_{KL}(r \parallel p_0)}{2}}, & \text{if } P_l \geq P_a, \\[8pt]
\dfrac{P_a}{1 - P_l + P_a} + \sqrt{\frac{E_{fit} + P_{uncov} D_{KL}(r \parallel p_0)}{2}}, & \text{if } P_l < P_a.
\end{cases}
\end{aligned}
\end{equation}

Thus, the second conclusion of the theorem is established, completing the proof.

Theorem 4 unifies the noise structure, coverage geometry, and uncovered-region strategies within a computable TVD upper bound framework. Specifically, the TVD upper bounds for model  $M $ with respect to both noisy actual inputs and true inputs are expressed in terms of the noise loss mass  $P_{l} $, addition mass  $P_{a} $, fitting error over the covered region  $E_{fit} $, uncovered mass  $P_{uncov} $, and the KL divergence $D_{KL}\left( r \parallel p_{0} \right) $ between the conditional distribution on the uncovered region and the prediction strategy. This formulation reveals several important insights: the loss and addition masses determine the deviation between the true target and the noisy target; the fit over the covered region, the coverage structure, and the divergence between the conditional distribution and prediction strategy on the uncovered region jointly determine the model\textquotesingle s deviation from the noisy target. Consequently, the effects of noise deletion/addition on target deviation can be directly assessed. This upper bound provides actionable guidance for the joint optimization of training coverage and prediction strategies, and can undoubtedly be applied more broadly to the analysis and optimization of model performance.

\section{Experimental Results}

Using the feedforward neural network model listed in Appendix C, we conducted regression experiments on two bivariate root functions: $y_{1} = \sqrt{4x_{1}^{2} + 3x_{2}^{2}} $   and
 $y_{2} = \sqrt{2x_{1}^{2} + x_{2}^{2}} $. Both the training and test datasets were specific. We evaluated the models' prediction errors under various conditions, assessed the impact and computational cost of ethical and safety mechanisms, and explored the challenge of providing multiple types of explanations for the same output. These experiments serve to illustrate and validate the definitions and theorems presented in Section 5.

\subsection{Experiments on the TVD upper bound under noiseless input and conservative prediction strategy}

To evaluate the TVD upper bound estimation of Theorem 3 under noiseless input and conservative prediction strategy, as well as its relationship to model performance, we construct the following experimental setup. On the  ${(x}_{1},x_{2}) $ plane, we select a square region centered at (1.0,1.0) with side length 1, and uniformly sample 10,000 points as the training dataset. For the test dataset, we uniformly sample 10,000 points within a circular region centered at (1.5,1.0) with radius 0.5. The model learning rate is set to  $\eta = 0.01 $. By uniformly sampling 10,000 points within the test region and computing the corresponding discrete points in the  ${(y}_{1},y_{2}) $ plane using the target function, we approximate the distribution  $p_{q}(x) $ of the test dataset and its entropy function. Fig. 3 illustrates the experimental conditions and outcomes:

(a) The overlap rate between the test and training datasets is approximately 50\%.

(b) For 10 randomly selected samples from the overlapping region, as the number of training steps increases from 10 to 4000 (i.e., as the model becomes more sufficiently trained), the root mean square error (RMSE) of the model's predictions steadily approaches zero. This result aligns with the first case of Theorem 3, equation (5.11).

{\sloppy\emergencystretch=1em
(c) For all test samples from the non-overlapping region, a conservative uniform prediction strategy was used. For almost all samples (except a few outliers), their RMSE is less than $\mathrm{TVD} \approx \sqrt{((1 - 0.4951)\ln{((10000 - 4951)}/(1 - 0.4951)) - 2.9741)/2} \approx 0.9155$. This indicates that even in regression tasks, although TVD is not strictly equivalent to generalization error, it can still effectively capture the distributional characteristics of generalization error.
\par}

(d) For 10 randomly selected test samples from the non-overlapping region, when using the model's prediction method, the RMSE also decreases with more training steps, reflecting a certain degree of model generalization outside the training data region. However, the rate of error reduction is significantly lower than in the overlapping region, and for some samples, the RMSE remains higher than that of the conservative uniform prediction. In fact, when training is insufficient, the overall RMSE of the model's predictions can even exceed the TVD. Meanwhile, the RMSE of conservative predictions for these samples is much lower than the TVD, reflecting the conclusion of the second case in Theorem 3, equation (5.11).

\begin{figure*}[ht]
\centering
\includegraphics[width=0.95\textwidth]{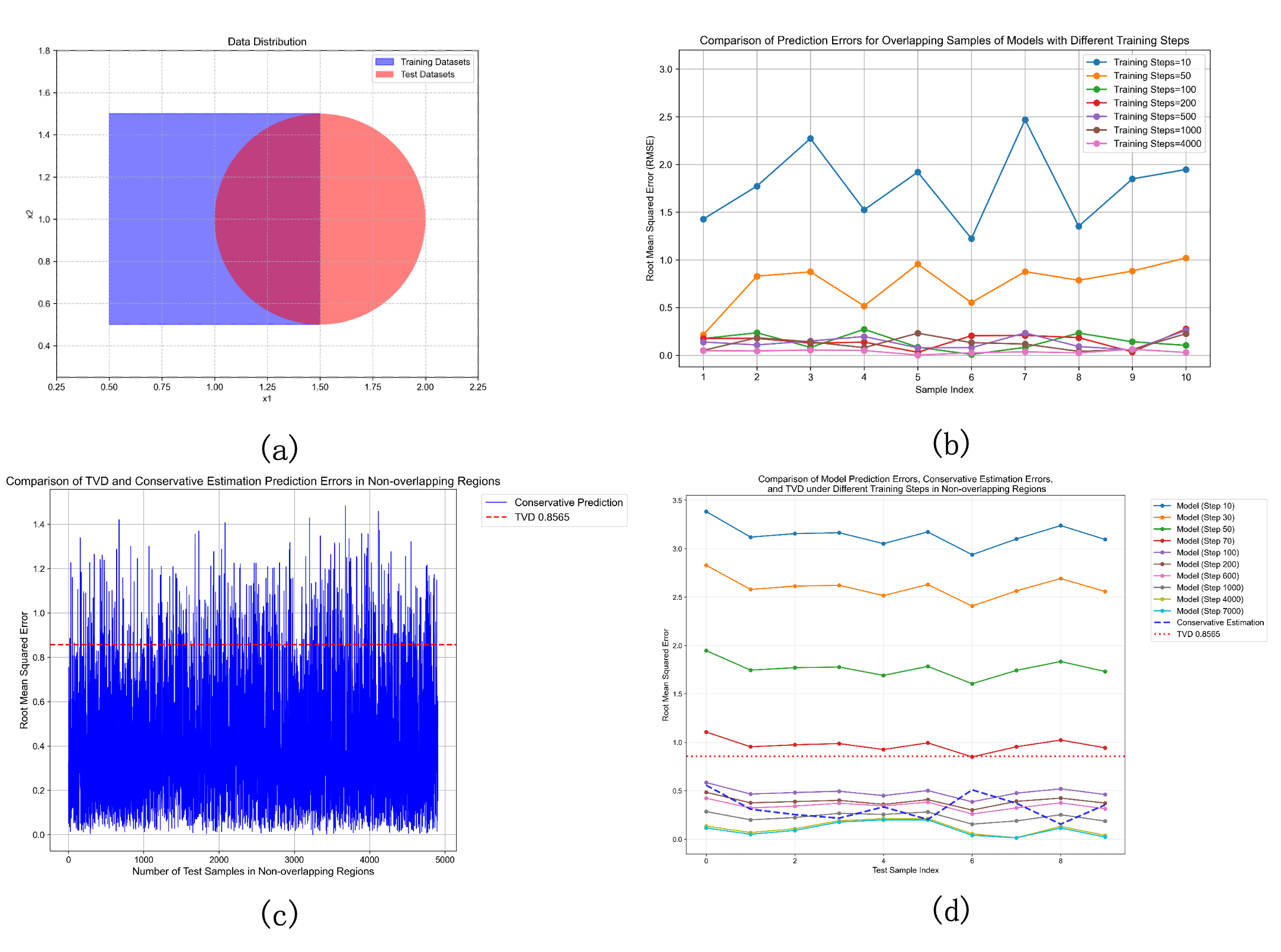}
\caption{Experimental Evaluation and Results of the TVD Upper Bound under Noiseless Input and Conservative Prediction Strategy.}
\label{fig:3}
\end{figure*}

\subsection{Experimental Evaluation of the TVD Upper Bound under Noisy Input and Arbitrary Prediction Strategy}

To evaluate Theorem 4's estimation of the TVD upper bound under conditions of noisy input and the adoption of arbitrary prediction strategies in regions not covered by the training data, as well as itsrelationship to model performance, we utilize the same model instance, training dataset, and test dataset as in the previous experiment. We introduce noise into the input by locally adding or deleting noise information in the test dataset according to Definition 12, so that the model input contains noise. Meanwhile, in the regions not covered by the
training data, we apply an arbitrary prediction strategy based on the joint distribution  $p_{0}(x) $ formed by two independent normal distributions,  $N\sim(0.6306,{0.4585}^{2}) $ and
 $N\sim\left( 0.4031,{0.3666}^{2} \right) $, to process and predict the input data. Fig. 4: (a) The training dataset  $S_{ou} $ (selected within the blue square) and the test dataset  $S_{oq} $ (selected within the red circle) are the same as those used in the experiment described in Section 6.1. The upper arc-shaped segment of  $S_{oq} $  is shifted upward, and the vacated region, denoted as  $S_{oq}^{-} $, is excluded from the test input. The shifted portion,  $S_{oq}^{+} $, does not overlap with the original test dataset but partially overlaps with  $S_{ou} $ ; it is added as new noisy test data. This structure satisfies the definition of noisy input, $\widetilde{S_{oq}} = S_{oq}\backslash S_{oq}^{-} \cup S_{oq}^{+} $.

\begin{figure*}[ht]
\centering
\includegraphics[width=0.95\textwidth]{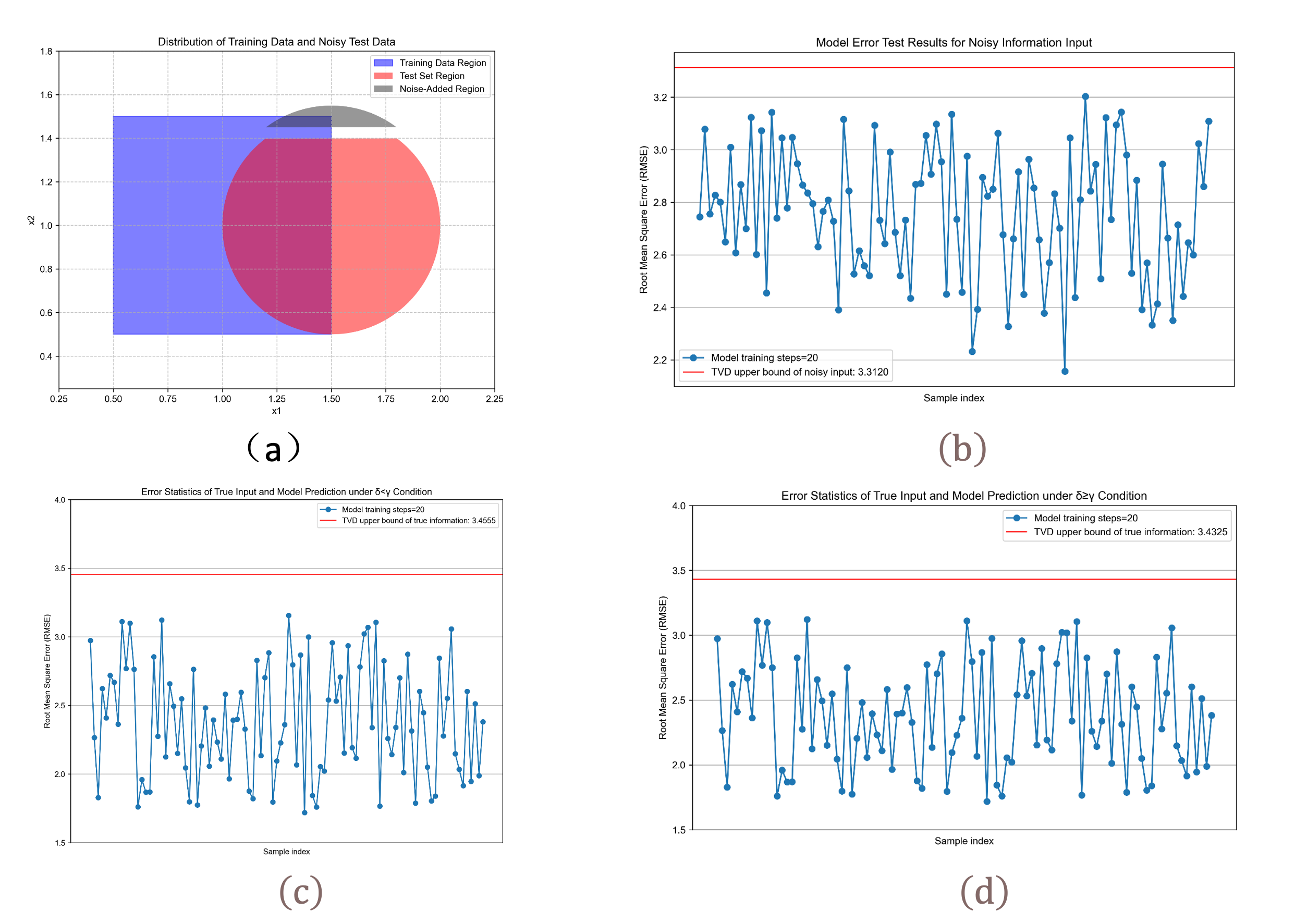}
\caption{Experimental Evaluation and Results of the TVD Upper Bound under Noisy Input and Arbitrary Prediction Strategy. }
\label{fig:4}
\end{figure*}

(b) For the noisy test dataset comprising 11,034 samples, we compute the probability distribution for each test sample using the same method as in Section 6.1. The loss mass is  $P_{l} = 0.0744 $, the addition mass is  $P_{a} = 0.0517 $, the uncovered mass is  $P_{uncov} = 0.5091 $, and the KL divergence is  $D_{KL}\left( r \parallel p_{0} \right) = 45.58 $. This yields a TVD upper bound between the noisy input distribution and the model distribution of 3.312. By randomly selecting 100 groups of samples from the test dataset and inputting them into the model after 20 training steps, we find that the root mean square error (RMSE) of all outputs (marked by blue dots) is less than the TVD upper bound indicated by the red line. This validates the conclusion of Theorem 4, Equation (5.17).

(c) In the above example,  $P_{l} = 0.0744 > 0.0517 = \gamma $. Under the same conditions, we randomly select 100 samples from the noiseless test dataset, convert them into corresponding noisy data as input to the model, and find that the RMSE of all outputs remains below the TVD upper bound of 3.4555. This validates the case where $P_{l} \geq P_{a} $  as given in Theorem 4, Equation (5.18).

(d) To examine the case where  $P_{l} < P_{a} $  in Theorem 4, Equation (5.18), we construct a new noisy test dataset comprising 10,000 samples. In this case,  $P_{l} = 0.0744 $,  $P_{a} = 0.1551 $, $P_{uncov} = 0.5530 $, and
 $D_{KL}\left( r \parallel p_{0} \right) = 40.88 $, resulting in a TVD upper bound of 3.4325. Again, by randomly selecting 100 samples from the noiseless test dataset and converting them into noisy inputs for the model, the RMSE of all outputs remains below the TVD upper bound indicated by the red line, thereby confirming the corresponding conclusion provided by Theorem 4.

Through these experiments, we have comprehensively validated Theorem 4 regarding the TVD upper bound under noisy input and arbitrary prediction strategies. Furthermore, as illustrated in Fig. 4, there remains a noticeable gap between the RMSE of the model outputs and the TVD upper bound across various scenarios, indicating that the upper bound provided
by Theorem 4 is still somewhat conservative. Further research may yield tighter error estimates under typical conditions.

\subsection{Ethical Safety Assurance Experiment}

Given the nature of the bivariate root function regression task, the model's predicted outputs should never be negative. This requirement is analogous to Definition 14, which ensures that model outputs do not violate the ethical constraints  $Ec $. Therefore, using the same training dataset as in the previous prediction error experiments, we set the learning rate to  $\eta = 0.001 $ and trained the model for 165 steps. For testing, we generated a dataset of 10,000 samples with  $(x_{1},x_{2}) $ drawn uniformly from within a circle of radius 0.5 centered at (-1, -1). The criterion for judging ethical safety violations was whether the model's predicted output was negative. This setup was used to test the feasibility of Theorem 2 and to evaluate the additional computational overhead introduced by ethical safety checking.

For each trial, 5,000 test samples were randomly selected using the trial number as a random seed. Two types of tests were conducted in parallel: ``Without Ethical Safety Check'' and ``With Ethical Safety Check.'' In the ``Without Ethical Safety Check'' condition, model predictions were output directly and the total computation time was recorded. In the ``With Ethical Safety Check'' condition, each prediction was first checked for sign; if any prediction was negative, an ethical error message was output, otherwise, the model's prediction was output. The total computation time was also recorded in this scenario. In total, 10 trials were conducted, and the results are shown in Table 1.

In Table 1, the ``Computation Time Ratio'' is calculated as the computation time with ethical safety checking divided by the computation time without it. For the simple feedforward neural network model used in these experiments, the additional computational overhead for performing ethical safety checks before model output ranged from 1.0\% to 19.2\%. As model size increases, this overhead is expected to decrease in proportion; however, as the number of ethical safety constraints grows, the overhead will increase accordingly. This is a key consideration for implementing ethical safety in models. In these 10 trials, more than 19\% of samples resulted in unethical predictions. In every trial, both the ``Correct Prediction Rate for Safe Outputs'' and the ``Correct Prompt Rate for Unethical Outputs'' were 100\%, demonstrating that by incorporating safety checks and informative outputs, we can always guarantee the ethical safety of model predictions while maximizing the number of safe outputs.

\begin{table}[htbp]
\centering
\caption{Summary Table of Ethical Safety Experiment Results}
\label{tab:ethical_safety}
\resizebox{\textwidth}{!}{%
{\renewcommand{\arraystretch}{1.2}
\setlength{\arrayrulewidth}{0.7pt}
\begin{tabular}{cccccccc}
\hline
Trial Number & 
\begin{tabular}[c]{@{}c@{}}Computation Time\\Without Ethical\\Safety Check\\(seconds)\end{tabular} & 
\begin{tabular}[c]{@{}c@{}}Computation Time\\With Ethical\\Safety Check\\(seconds)\end{tabular} & 
\begin{tabular}[c]{@{}c@{}}Computation\\Time Ratio\end{tabular} & 
\begin{tabular}[c]{@{}c@{}}Number of\\Ethically Safe\\Output Samples\end{tabular} & 
\begin{tabular}[c]{@{}c@{}}Number of\\Unethical\\Warning Samples\end{tabular} & 
\begin{tabular}[c]{@{}c@{}}Correct Prediction\\Rate for Safe\\Outputs\end{tabular} & 
\begin{tabular}[c]{@{}c@{}}Correct Prompt\\Rate for\\Unethical Outputs\end{tabular} \\
\hline
1  & 0.367 & 0.391 & 1.063 & 4517 & 483 & 100\% & 100\% \\
2  & 0.365 & 0.378 & 1.036 & 4521 & 479 & 100\% & 100\% \\
3  & 0.352 & 0.362 & 1.030 & 4548 & 452 & 100\% & 100\% \\
4  & 0.330 & 0.378 & 1.146 & 4516 & 484 & 100\% & 100\% \\
5  & 0.371 & 0.383 & 1.032 & 4523 & 477 & 100\% & 100\% \\
6  & 0.327 & 0.380 & 1.160 & 4546 & 454 & 100\% & 100\% \\
7  & 0.358 & 0.402 & 1.123 & 4545 & 455 & 100\% & 100\% \\
8  & 0.360 & 0.410 & 1.140 & 4538 & 462 & 100\% & 100\% \\
9  & 0.361 & 0.365 & 1.010 & 4519 & 481 & 100\% & 100\% \\
10 & 0.355 & 0.424 & 1.192 & 4533 & 467 & 100\% & 100\% \\
\hline
\end{tabular}}%
}
\end{table}

\subsection{Interpretability Experiment}

For our bivariate root regression task, the output should never be negative. However, in the ethical safety assurance experiment, after 165 training steps, the model produced negative predictions for more than 19\% of randomly selected 5,000-sample test batches, repeated over 10 trials. According to Theorem 1, this phenomenon can be explained from multiple perspectives, each based on different information.

\subsubsection{Mathematical Operation-Based Explanation}

After 165 training steps, the neural network model M has the following weight matrices and bias vectors from the input layer to the hidden layer:

\begin{equation}
W_{1} \approx \begin{bmatrix}
-0.065 & -0.429 \\
0.681 & -0.157 \\
-0.498 & -0.586
\end{bmatrix}, \quad b_{1} \approx \begin{bmatrix}
0.129 \\
-0.551 \\
0.350
\end{bmatrix}
\label{eq:weights_bias_1}
\end{equation}

The weights and biases from the hidden layer to the output layer are:

\begin{equation}
W_{2} \approx \begin{bmatrix}
0.145 & 0.519 & -0.423 \\
0.022 & 0.048 & 0.043
\end{bmatrix}, \quad b_{2} \approx \begin{bmatrix}
0.613 \\
0.199
\end{bmatrix}
\label{eq:weights_bias_2}
\end{equation}

Therefore, for the test sample $X = [-1.3535, -1.3535]^{T}$, the model predicts:

\begin{equation}
Y = \begin{bmatrix}
y_{1} \\
y_{2}
\end{bmatrix} = \text{Identity}\left( W_{2}\text{ReLU}\left( W_{1}X + b_{1} \right) + b_{2} \right) \approx \begin{bmatrix}
-0.040 \\
0.277
\end{bmatrix}
\label{eq:prediction_result}
\end{equation}

Here,  $y_{1} = - 0.040 < 0 $, which mathematically explains why model M outputs a negative value for input  $X $.

These formulas provide the designer's mathematical representation of the model's computational process. They are equivalent to the formal forward propagation rules (C.21) and (C.22) in Appendix C. According to the OIT perspective, these equations describe the ontological state of the model, its outputs, and the information they convey. The model, along with its input and output mechanisms, serves as the carrier of this information and realizes the computational process. Therefore, using these formulas to explain why the model produces negative outputs
satisfies the requirements of Definition 13 regarding model interpretability and meets the needs of researchers who seek to understand model behavior from a mathematical standpoint.

\subsubsection{Visualization-Based Explanation of the Computational Process}
Depicting the model's computational process visually can make explanations of model behavior much more intuitive. To this end, for model M after 165 training steps, we illustrate its complete structure---covering the inputs, internal neural network calculations and data flow, and outputs---in Fig. 5.

\begin{figure*}[ht]
\centering
\includegraphics[width=0.95\textwidth]{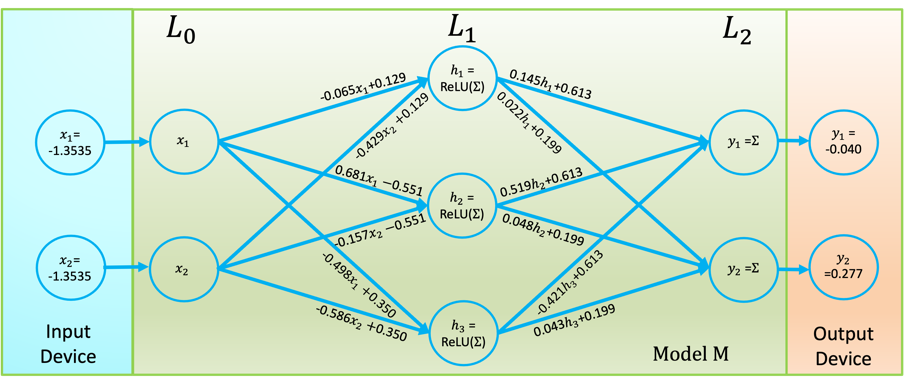}
\caption{Schematic diagram of the computational process for model M after 165 training steps.}
\label{fig:5}
\end{figure*}

Compared with equations \eqref{eq:weights_bias_1}--\eqref{eq:prediction_result}, Fig. 5 not only comprehensively reflects the input, weight matrices, bias vectors, neuron activation functions, and the specific data and formulas for outputs, but also highlights the interactions between the input/output devices and the model. The figure offers a more direct view of the internal data processing pipeline, making it easier to examine and verify each step of the computation. Thus, the visualization provides a more accessible and concrete explanation of how model M processes inputs and generates results.

\subsubsection{Partition-Based Explanation}

It is evident that equations \eqref{eq:weights_bias_1}--\eqref{eq:prediction_result} and Fig. 5 explain only the result of model M processing the specific sample $X = [-1.3535, -1.3535]^{T}$. They do not account for the phenomenon observed in the ethical safety assurance experiment, where each random selection of 5,000 samples from the test dataset yields over 19\% negative predictions. To explain this, a broader statistical analysis across more data samples is needed.

Fig. 6(a) provides a visual depiction of the decision boundary after 165 training steps, showing where model M predicts negative outputs. Notably, there is a boundary line approximately passing through points (-2.0, -0.4) and (-0.4, -2.0). For samples on the upper right side of this diagonal, model M consistently predicts non-negative values; for those on the lower left, the model consistently outputs negative values. The red region in the figure, representing the test dataset, falls precisely on the lower left side of this boundary line, which explains why a random selection of 5,000 samples from the test set will always include a substantial portion that triggers negative predictions. This visual and statistical partitioning thus provides an intuitive explanation for the observed frequency of negative outputs in the ethical safety experiment.

\begin{figure*}[ht]
\centering
\includegraphics[width=0.95\textwidth]{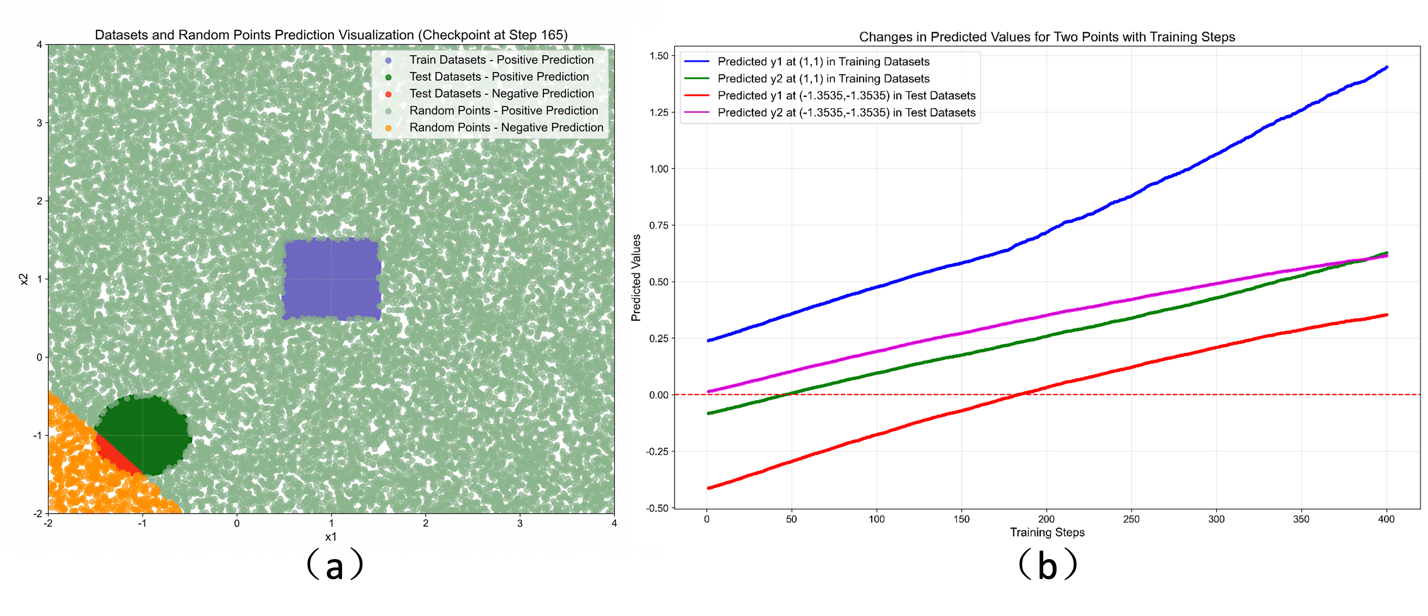}
\caption{Illustration of Prediction Polarity and Impact of Training Steps.}
\label{fig:6}
\end{figure*}

\subsubsection{Explanation Based on Cumulative Training Effects}

The previous three explanations all focus on the behavior of M after 165 training steps. However, model training is a continuous process of cumulative improvement. Examining only the model\textquotesingle s instantaneous performance at a particular training step cannot fully explain the causes of abnormal outputs. To address this, we further trained M for 400 iterations on the training dataset. Fig. 6(b) shows the predictions for both (1, 1) and (-1.3535, -1.3535) after each training step.

It is clear that even for (1, 1) --- a central sample from the training set---the predicted  $y_{2} $ by M remains negative during roughly the first 50 training steps. Only after more than 50 steps does the prediction move away from negative values. For the test sample (-1.3535, -1.3535), it takes more than about 180 training steps before the prediction begins to consistently avoid negative values. This cumulative effect is entirely due to the optimization of M's parameters as training progresses.

In fact, by applying the forward propagation rules (C.21) and (C.22) from the appendix---substituting the model parameters obtained after each training step---one can, through a recursive process, provide a rigorous and step-by-step explanation for M's predictions on specific input samples after any given amount of training.

\section{Conclusion}

This work, grounded in the Objective Information Theory (OIT), develops the MLT-MF framework, employing a formal logical system to integrate the complete causal chain of model construction, training, input processing, output, and online learning. Under a unified semantics, the author proposes formal definitions of model interpretability and ethical safety, derive two classes of explicit TVD upper bound theorems, and validate the framework through experiments with feedforward neural network models. 

Of course, this study has limitations: the current formal system is most transparent for finite discrete sets, and further refinement is needed for continuous, time-varying, multimodal, and multi-agent scenarios, particularly regarding measurability and dynamic consistency. The maximal independent set for ethical safety is generally NP-hard, requiring the development of complexity-controlled approximation algorithms and alignment with business ethics standards. TVD upper bounds rely on conservative inequalities such as Pinsker’s; in typical scenarios, error estimates may be tightened using related theoretical advances. Physical causality in enabled mappings calls for interdisciplinary integration with various branches of physics, leveraging parameters such as energy consumption, latency, and uncertainty to advance from qualitative causality to quantitative verification. Future work may include large-scale empirical studies to verify the observability of TVD upper bounds and the benefits of strategy optimization, as well as further development of cross-modal and multi-agent information chain composition and conflict resolution methods, thereby enhancing the universality and practical value of MLT-MF in complex systems.


\section*{Acknowledgement}
Junxuan He from Beijing University of Posts and Telecommunications implemented all the experiments described in the paper according to the author’s ideas, producing complete model code and training/test datasets, providing indispensable and strong support for illustrating and validating the theorems related to MLT-MF. The author offers special thanks for his hard work, keen understanding, and efficient implementation. During the writing and revision of this paper, I received many insightful comments from Associate Professor Wang Rui of the School of Computer Science at Shanghai Jiao Tong University. I also gained much inspiration and assistance from regular academic discussions with doctoral students Siyuan Qiu, Chun Li, Hu Xu, and Zeyan Li. I hereby express my sincere gratitude to them.

\section*{Declaration of competing interest}
The author declare that he has no known competing ﬁnancial interests or personal relationships that could have appeared to inﬂuence the work reported in this paper.

\section*{Data and Code availability}
{\sloppy\emergencystretch=1em
All datasets and codes used in this work are available at https://github.com/xuhu6736/MLT-MF.
\par}


\bibliographystyle{unsrt}
\bibliography{ref}

\newpage

\appendix

\section{ Formal Representation of States}

The postulates of information emphasize that the state of the ontology enables a mapping to the state of the carrier. Therefore, establishing a formal representation of both ontological and carrier states is particularly important.

\textbf{Definition A.1 (Higher-Order Formal Logical System)}: Suppose a formal system  $\mathcal{L} $ is described using the following symbol set:

Variables  $x_{1} $,  $x_{2} $,\ldots;

Individual constants  $a_{1} $,  $a_{2} $,\ldots;

Functions
 $f_{1}^{1} $,  $f_{2}^{1} $, \ldots{} $f_{1}^{2} $,  $f_{2}^{2} $, \ldots{} $f_{1}^{3} $,  $f_{2}^{3} $, \ldots;

Parentheses  $(,) $;

Predicates
 $A_{1}^{1} $,  $A_{2}^{1} $, \ldots{} $A_{1}^{2} $,  $A_{2}^{2} $, \ldots{} $A_{1}^{3} $,  $A_{2}^{3} $, \ldots;

Logical connectives ``not''  $\sim $, ``implication'' $\rightarrow $;

Quantifier ``for all''  $\forall $.

A term in  $\mathcal{L} $ is defined as follows:

(1) Variables and individual constants are terms;

(2) If  $f_{i}^{n}(n > 0,i > 0) $ is a function in  $\mathcal{L} $, and  $t_{1} $,\ldots,  $t_{n} $ are terms in  $\mathcal{L} $ that do not contain  $f_{i}^{n} $ itself, then  $f_{i}^{n}(t_{1} $,\ldots,  $t_{n}) $ is also a term in  $\mathcal{L} $;

(3) If  $A_{i}^{n}(n > 0,i > 0) $ is a predicate in  $\mathcal{L} $, and  $t_{1} $,\ldots,  $t_{n} $ are terms in  $\mathcal{L} $ that do not contain  $A_{i}^{n} $ itself, then  $A_{i}^{n}(t_{1} $,\ldots,  $t_{n}) $ is also a term in  $\mathcal{L} $;

(4) The set of all terms in  $\mathcal{L} $ is generated by (1), (2),
and (3).

Based on this, we can define atomic formulas in  $\mathcal{L} $:

If  $A_{i}^{n}(n > 0,i > 0) $ is a predicate in  $\mathcal{L} $ and  $t_{1} $,\ldots,  $t_{n} $ are terms in  $\mathcal{L} $ not containing $A_{i}^{n} $ itself, then  $A_{i}^{n}(t_{1} $,\ldots,  $t_{n}) $ is an atomic formula in  $\mathcal{L} $.

We then define well-formed formulas (WFFs) in  $\mathcal{L} $:

(5) Every atomic formula in  $\mathcal{L} $ is a WFF;

(6) If  $\mathcal{A} $ and  $\mathcal{B} $ are WFFs in  $\mathcal{L} $,then  $\mathcal{\sim A} $,  $\mathcal{A \rightarrow B} $,  $\ (\forall x_{i}\mathcal{)A} $,  $(\forall f_{i}^{n}\mathcal{)A} $, and  $(\forall A_{i}^{n}\mathcal{)A} $ are all WFFs in  $\mathcal{L} $, where  $x_{i} $,  $f_{i}^{n} $,  $A_{i}^{n} $ are variables, functions, and predicates in  $\mathcal{L} $, respectively, and no predicate or function symbol is applied, either directly or indirectly, to itself;

(7) The set of all WFFs in  $\mathcal{L} $ is generated by (5) and (6).

The higher-order formal logical system  $\mathcal{L} $ defined here, based on the flexibility of predicate logic, can describe various relationships and constraints among states. Its recursive structure and higher-order quantification support the nested definition of complex data and behaviors, and it explicitly prohibits any predicate or function symbol from applying to itself---directly or indirectly---to avoid self-referential paradoxes.

For simplicity,  $\mathcal{L} $ does not define connectives such as ``and''  $\land $, ``or''  $\vee $, ``if and only if''  $\leftrightarrow $, or the quantifier ``exists''  $\exists $, and other commonly used logical symbols. However, mathematical logic theory demonstrates that these symbols are entirely equivalent to appropriate combinations of existing symbols and can make many formulas more concise \citep{hamilton1988logic}. Therefore, they are introduced as defined symbols in subsequent discussions.

\textbf{Definition A.2 (Interpretation of a Formal System)} An interpretation  $E $ of a formal system  $\mathcal{L} $ over a domain  $D_{E} $ assigns specific values to constants, elements of the domain  $D_{E} $ to variables, and turns function and predicate symbols into functions and predicates over that domain  $D_{E} $, thereby yielding a set of WFFs each with a truth value.

Definition A.2 assigns concrete values and meaning to a set of formulas in a formal system  $\mathcal{L} $ through interpretation, thereby providing a fundamental approach for formally describing the definitional, attributive, structural, relational, and evolutionary state properties of objects in both the subjective and objective worlds.

\textbf{Postulate A.1 (Expression of Object States)} The state set of an object collection over a specified time set is constituted by its attributes---such as properties, structures, values, and relations---as well as the evolution rules and logical combinations among these attributes, and satisfies the following conditions:

(1) \textbf{Domain invariance:} definitions of the object set and time set remain unchanged.

(2) \textbf{Numerical determinacy:} concrete numerical values determine values of certain properties and forms.

(3) \textbf{Attribute expressibility:} properties, forms, values, relations are expressible via functions and predicates in a formal system satisfying Definition 2.

(4) \textbf{Rule expressibility:} evolution rules among attributes are expressible in such a formal system using time variables and implication, and contain no implicit self-reference.

(5) \textbf{Logical coherence:} no logical contradictions exist among attributes and rules, nor among any logical conjunctions thereof.

\textbf{Lemma A.1 (Formal Expression of States)} If the state set $S(X,T) $ of objects  $X $ over times  $T $ satisfies Postulate A.1, then it can be represented as an interpretation of a formal system $\mathcal{L} $ over the domain  $X \times T $, and it is logically coherent.

\textbf{Proof}: Since the state set  $S(X,T) $ of the object set  $X $ over the time set  $T $ satisfies Postulate A.1, we construct the corresponding formal system  $\mathcal{L} $ as follows:

By (1), the definitions of the elements in  $X $ and  $T $ remain unchanged, ensuring determinacy. The variables in the formal system $\mathcal{L} $ are defined to correspond to each element of  $X \times T $, thus making  $X \times T $ the domain of discourse for  $\mathcal{L} $. Consequently,  $S(X,T) $ has a well-defined domain.

By (2), the constants in  $\mathcal{L} $ are defined to represent all possible values in  $S(X,T) $.

By (3), the functions and predicates in  $\mathcal{L} $ represent the properties, forms, values, and relations of  $X $ over  $T $ as captured in  $S(X,T) $.

{\sloppy\emergencystretch=1em
By (4), consider two ternary predicates  $A_{1}^{3} $  and $A_{2}^{3} $. Let  $A_{1}^{3}(x_{1},t_{1},f_{1}^{2}\left( x_{1},t_{1} \right)) $ and  $A_{2}^{3}(x_{2},t_{2},f_{2}^{2}\left( x_{2},t_{2} \right)) $ be two properties of  $S(X,T) $ corresponding to objects  $x_{1},x_{2} \in X $, times  $t_{1},t_{2} \in T(t_{1} < t_{2}) $, and binary functions $f_{1}\left( x_{1},t_{1} \right),f_{2}\left( x_{1},t_{1} \right) $. Suppose all $A_{1}^{3}(x_{1},t_{1},f_{1}^{2}\left( x_{1},t_{1} \right)) $ satisfy the rule of evolving from  $t_{1} $  to  $t_{2} $ into some $A_{2}^{3}(x_{2},t_{2},f_{2}^{2}\left( x_{2},t_{2} \right)) $. The formula $(\forall A_{1}^{3}\forall f_{1}^{2}\forall x_{1}A_{1}^{3}(x_{1},t_{1},f_{1}^{2}\left( x_{1},t_{1} \right)) \rightarrow {\exists A_{2}^{3}\exists f_{2}^{2}\exists x_{2}A}_{2}^{3}(x_{2},t_{2},f_{2}^{2}\left( x_{2},t_{2} \right)) $ is also a well-formed formula in  $\mathcal{L} $, with respect to the interpretation of  $x_{1},x_{2},t_{1},t_{2} $, and is able to express the corresponding rules in  $S(X,T) $. For other rules in  $S(X,T) $, the related elements in the domain  $X \times T $ can similarly be interpreted and expressed via corresponding WFFs in $\mathcal{L} $, with no implicit logical self-reference.
\par}

Therefore,  $S(X,T) $ constitutes an interpretation  $E $ of  $\mathcal{L} $ over the domain  $X \times T $. By (5), both itself and its logical closure are logically consistent. The lemma is thus proved.

Lemma A.1 stipulates that the variables in  $\mathcal{L} $ are elements of  $X \times T $; thus, in formulas, object and time elements appear in pairs  $(x,t) $. In practice, some attributes may be time-invariant; then simplified expressions may be used to describe such attributes without time.

\section{ Finite Automaton States}

Using the formal expression of states, we can study in depth the special and important object of finite automata \citep{editorial2002dictionary}.

\textbf{Lemma B.1 (State Representation of Finite Automata)} For any finite automaton  $M $, one can describe its input, output, and state‐transition behavior over a related time set  $T $ by the state set
 $S(M,T) $.

{\sloppy\emergencystretch=1em
\textbf{Proof:} Let
 $M = \left\langle Q,R,U,\delta,\lambda \right\rangle $ be a finite automaton, and  $T = \left\{ t_{i} \middle| i = 1,\ldots,n \right\} $ the (strictly increasing) time points at which  $M $ performs input, output, or state‐transition actions. By finite‐automaton theory \citep{editorial2002dictionary},  $Q = \left\{ q_{t_{i}} \middle| i = 1,\ldots,n - 1 \right\} $, $R = \left\{ r_{t_{i}} \middle| i = 2,\ldots,n \right\} $, and  $U = \left\{ u_{t_{i}} \middle| i = 1,\ldots,n \right\} $ are nonempty finite sets of input symbols, output symbols, and states, respectively;
 \par}

 $\delta $:  $U\mathbf{\times}Q $ to  $U $, is the next‐state function;

 $\lambda $:  $U\mathbf{\times}Q $ to  $R $, is the output function.

For each state  $u_{t_{i}} \in U $, define the state predicate
 ${State}^{3} $ by
\begin{equation}
\varphi_{U}(t_{i}) = {State}^{3}(M,t_{i},u_{t_{i}}) \tag{B.1}
\end{equation}
to assert that at time  $t_{i} $,  $M $ is in state  $u_{t_{i}} $,
 $i = 1,\ldots,n $.

For each input  $q_{t_{i}} \in Q $, define the input predicate
 ${Input}^{3} $ by
\begin{equation}
\varphi_{I}(t_{i}) = {Input}^{3}(M,t_{i},q_{t_{i}}) \tag{B.2}
\end{equation}
to assert that at time  $t_{i} $,  $M $ receives input  $q_{t_{i}} $,
 $i = 1,\ldots,n - 1 $.

For each output  $r_{t_{i}} \in R $, define the output predicate ${Output}^{3} $ by
\begin{equation}
\varphi_{O}(t_{i}) = {Output}^{3}(M,t_{i},r_{t_{i}}) \tag{B.3}
\end{equation}
to assert that at time  $t_{i} $,  $M $ produces output  $r_{t_{i}} $,  $i = 2,\ldots,n $.

The transition function  $\delta $ is captured by the well‐formed formula
\begin{equation}
\varphi_{\delta}\left( t_{i} \right) = \varphi_{U}(t_{i}) \land \varphi_{I}(t_{i}) \rightarrow \exists\delta(u_{t_{i}},q_{t_{i}}) \land {Eq}^{2}(u_{t_{i + 1}},\delta(u_{t_{i}},q_{t_{i}})) \land \varphi_{U}(t_{i + 1}) \tag{B.4}
\end{equation}
where the binary predicate  ${Eq}^{2}(x,y) $ asserts  $x = y $,
 $i = 1,\ldots,n - 1 $.

Similarly, the output function  $\lambda $ is given by
\begin{equation}
\begin{aligned}
\varphi_{\lambda}\left( t_{i} \right) = &\ \varphi_{U}(t_{i}) \land \varphi_{I}(t_{i}) \rightarrow \exists\lambda(u_{t_{i}},q_{t_{i}}) \land \\
& \mathit{Eq}^{2}(r_{t_{i + 1}},\lambda(u_{t_{i}},q_{t_{i}})) \land \\
& \mathit{IsElementof}^{2}(r_{t_{i + 1}},R) \land \varphi_{O}(t_{i + 1})
\end{aligned}
\tag{B.5}
\end{equation}
where  ${IsElementof}^{2}(x,X) $ asserts  $x $ is an element of  $X $,
 $i = 1,\ldots,n - 1 $.

Thus the state set
\begin{equation}
\begin{split}
S(M,T) = \{& \varphi_{U}\left( t_{i} \right),\varphi_{I}\left( t_{j} \right),\varphi_{O}\left( t_{k} \right),\varphi_{\delta}\left( t_{j} \right),\varphi_{\lambda}\left( t_{j} \right), \\
& i = 1,\ldots,n,j = 1,\ldots,n - 1,k = 2,\ldots,n \}
\end{split} \tag{B.6}
\end{equation}
describes  $M $'s inputs, outputs, and state‐transitions over  $T $. Here  $Cn(X) $ denotes the logical closure of the set of WFFs  $X $. This completes the proof.

It follows that every finite automaton can be formalized by the state set given in Theorem 2, which moreover exhibits automatic transition
behavior. Hence we introduce:

\textbf{Definition B.2 (Finite Automaton State)} If a state set $S(X,T) $ expresses a finite automaton, then  $S(X,T) $ is called a finite automaton state.

All practical information systems, including computing systems, are finite automata in the mathematical sense. Therefore, the concept of finite automaton state is of great significance in the study of machine learning.

\section{ Model Instance Validation of MLT-MF}
\setcounter{figure}{0} 
\setcounter{table}{0} 

To facilitate validation and illustration of the practical significance of MLT-MF and related theorems, we select a very simple model and use MLT-MF to establish formal representations with WFFs for each stage from initial information to training information, learning optimization, and processing/output.

\textbf{C.1 A Simple Feedforward Neural Network Model}

We choose a feedforward neural network M with an input layer containing two inputs, a hidden layer containing three neurons, and an output layer containing two outputs (Fig. C.1). Its simplicity helps explain the essence of MLT-MF in a clear and straightforward manner.
\begin{figure*}[ht]
\centering
\includegraphics[width=0.75\textwidth]{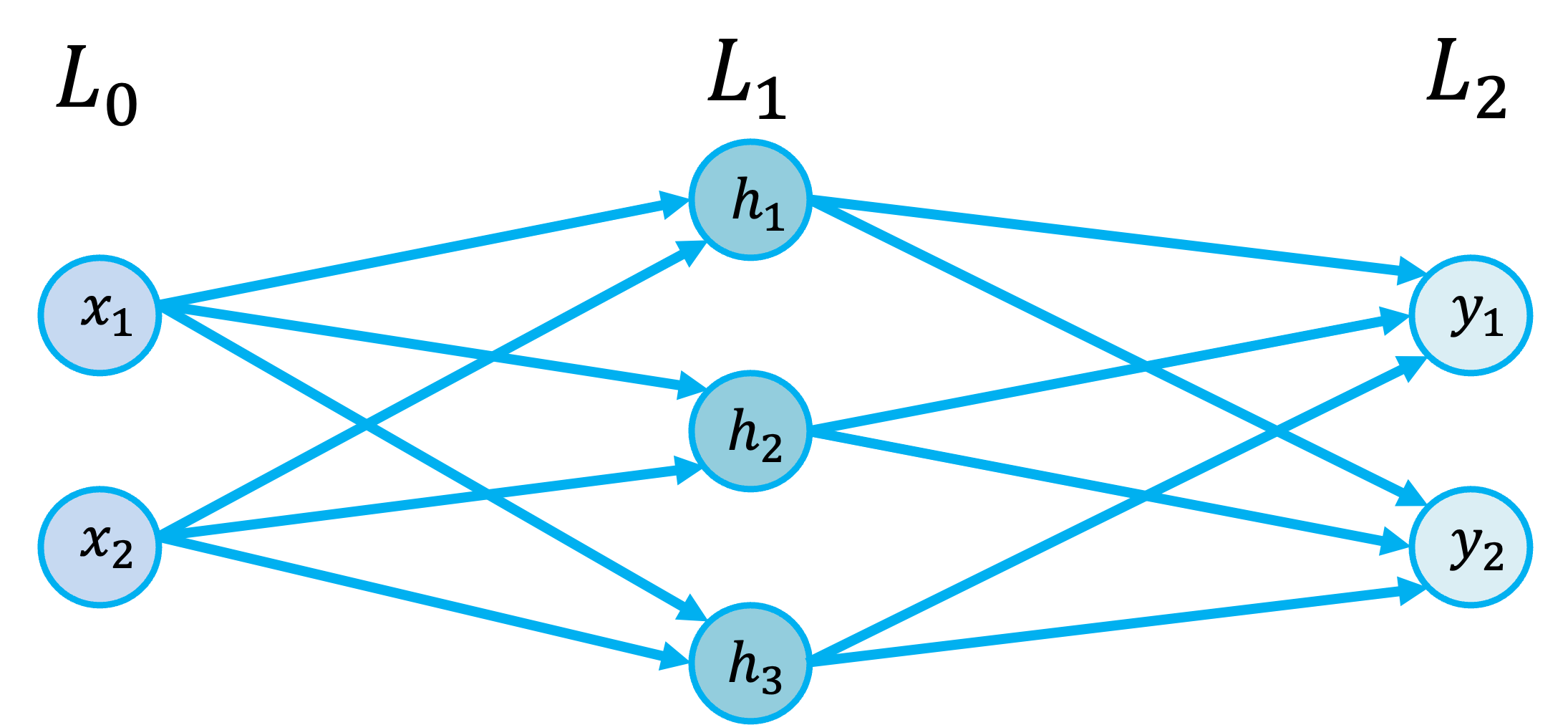}
\caption{Schematic diagram of feedforward neural network model M}
\label{}
\end{figure*}

The hidden layer uses the ReLU activation function:
\begin{equation}
ReLU(x) = max(0,x)
\end{equation}

The output layer uses the identity activation:
\begin{equation}
Identity(x) = x
\end{equation}

Initial weight matrices and bias vectors are defined between layers:
\begin{equation}
W_{1} = \begin{bmatrix}
w_{11}^{1} & w_{12}^{1} \\
w_{21}^{1} & w_{22}^{1} \\
w_{31}^{1} & w_{32}^{1}
\end{bmatrix}, \quad b_{1} = \begin{bmatrix}
b_{1}^{1} \\
b_{2}^{1} \\
b_{3}^{1}
\end{bmatrix}
\end{equation}
\begin{equation}
W_{2} = \begin{bmatrix}
w_{11}^{2} & w_{12}^{2} & w_{13}^{2} \\
w_{21}^{2} & w_{22}^{2} & w_{23}^{2}
\end{bmatrix}, \quad b_{2} = \begin{bmatrix}
b_{1}^{2} \\
b_{2}^{2}
\end{bmatrix}
\end{equation}

Training or input data, output data, and training labels or target outputs are specified, with mean squared error (MSE) as the loss:
\begin{equation}
X = \begin{bmatrix}
x_{1} \\
x_{2}
\end{bmatrix}, \quad Y = \begin{bmatrix}
y_{1} \\
y_{2}
\end{bmatrix}, \quad Y_{t} = \begin{bmatrix}
y_{t1} \\
y_{t2}
\end{bmatrix} \text{ and } Y_{t} = Target(X)
\end{equation}
\begin{equation}
L = MSE\left( Y, Y_{t} \right) = \frac{1}{2}\left\| Y - Y_{t} \right\|^{2} = \frac{1}{2}(\left( y_{1} - y_{t1} \right)^{2} + \left( y_{2} - y_{t2} \right)^{2})
\end{equation}

The gradients of the loss function  $L $ with respect to the weight matrices and bias vectors are the primary basis for model M's optimization during training. The four calculation formulas are as follows:
\begin{equation}
\nabla_{W_{2}}L = (Y - Y_{t}){(ReLU(W_{1}X + b_{1}))}^{T}
\end{equation}
\begin{equation}
\nabla_{b_{2}}L = (Y - Y_{t})
\end{equation}
\begin{equation}
\nabla_{W_{1}}L = ({W_{2}}^{T}(Y - Y_{t})\bigodot{ReLU}'(W_{1}X + b_{1}))X^{T}
\end{equation}
\begin{equation}
\nabla_{b_{1}}L = {W_{2}}^{T}(Y - Y_{t})\bigodot{ReLU}'(W_{1}X + b_{1})
\end{equation}

In equations (C.9) and (C.10), the operator  $\bigodot $ denotes element-wise multiplication, and  ${ReLU}' $ represents the derivative of the  $ReLU $ function. Since  $ReLU $ is not differentiable at zero, we additionally define  ${ReLU}'(0) = 0 $. Thus, the update formulas for optimizing the weight matrices and bias vectors during M's training are:
\begin{equation}
W_{1}^{\text{new}} = W_{1} - \eta\nabla_{W_{1}}L, \quad b_{1}^{\text{new}} = b_{1} - \eta\nabla_{b_{1}}L
\end{equation}
\begin{equation}
W_{2}^{\text{new}} = W_{2} - \eta\nabla_{W_{2}}L, \quad b_{2}^{\text{new}} = b_{2} - \eta\nabla_{b_{2}}L
\end{equation}
where  $\eta $ is the learning rate of M.

\textbf{C.2 Symbol Definitions}

MLT-MF uses a formal system to express the operational logic of each phase throughout the model's full lifecycle. Defining the symbols is a prerequisite for formal modeling. To this end, Tables C.1--4 provide definitions for the constants, variables, functions, and predicates used in the formal representation of model M.

\begin{table}[H]
\centering
\caption{Definition of Constants}
\label{tab:constants}
{\renewcommand{\arraystretch}{1.2}
\setlength{\arrayrulewidth}{0.7pt}
\begin{tabular}{p{0.2\textwidth}p{0.2\textwidth}p{0.5\textwidth}}
\hline
Symbol & Type & Definition \\
\hline
$\mathbb{R}^2$ & Space & 2-dimensional real space \\
$\mathbb{R}^3$ & Space & 3-dimensional real space \\
\hline
\end{tabular}}
\end{table}

{\renewcommand{\arraystretch}{1.2}
\setlength{\arrayrulewidth}{0.7pt}
\begin{longtable}{p{0.15\textwidth}p{0.25\textwidth}p{0.55\textwidth}}
\caption{Definition of Variables} \label{tab:variables} \\
\hline
Symbol & Type & Definition \\
\hline
\endfirsthead

\multicolumn{3}{c}{\tablename\ \thetable{} -- continued from previous page} \\
\hline
Symbol & Type & Definition \\
\hline
\endhead

\hline
\multicolumn{3}{r}{Continued on next page} \\
\endfoot

\hline
\endlastfoot

$i,j,k$ & natural number & index or count \\
$\varphi_i^o$ & WFFs & WFFs contained in the informational ontological state of M's initial model \\
$\psi_i^o$ & WFFs & WFFs contained in the informational carrier state of M's initial model \\
$\varphi_i^t$ & WFFs & WFFs contained in the training informational ontological state \\
$\psi_i^t$ & WFFs & WFFs contained in the training informational carrier state \\
$\varphi_i^u$ & WFFs & WFFs contained in the informational ontological state of M's optimized model \\
$\psi_i^u$ & WFFs & WFFs contained in the informational carrier state of M's optimized model \\
$\varphi_i^q$ & WFFs & WFFs contained in the input informational ontological state of M \\
$\psi_i^q$ & WFFs & WFFs contained in the input informational carrier state of M \\
$\varphi_i^r$ & WFFs & WFFs contained in the output informational ontological state of M \\
$\psi_i^r$ & WFFs & WFFs contained in the output informational carrier state of M \\
$\varphi_i^v$ & WFFs & WFFs contained in the informational ontological state of M's re-optimized model \\
$\psi_i^v$ & WFFs & WFFs contained in the informational carrier state of M's re-optimized model \\
layers & natural number & number of layers in the model \\
$v$ & real number & numerical values for weights and biases \\
$\mathbf{W}_k^{(o)}$ & matrix & weight matrix from the $k$-th to the $(k+1)$-th layer in the initial model \\
$w_{ij}^{(o)k}$ & real number & weight from the $j$-th neuron in the $k$-th layer to the $i$-th neuron in the $(k+1)$-th layer in the initial model \\
$\mathbf{b}_k^{(o)}$ & vector & bias vector of neurons in the $k$-th layer in the initial model \\
$b_i^{(o)k}$ & real number & bias of the $i$-th neuron in the $k$-th layer in the initial model \\
$\mathbf{W}_k^{(u)}$ & matrix & weight matrix from the $k$-th to the $(k+1)$-th layer in the optimized model \\
$w_{ij}^{(u)k}$ & real number & weight from the $j$-th neuron in the $k$-th layer to the $i$-th neuron in the $(k+1)$-th layer in the optimized model \\
$\mathbf{b}_k^{(u)}$ & vector & bias vector of neurons in the $k$-th layer in the optimized model \\
$b_i^{(u)k}$ & real number & bias of the $i$-th neuron in the $k$-th layer in the optimized model \\
$\mathbf{W}_k^{(v)}$ & matrix & weight matrix from the $k$-th to the $(k+1)$-th layer in the re-optimized model \\
$w_{ij}^{(v)k}$ & real number & weight from the $j$-th neuron in the $k$-th layer to the $i$-th neuron in the $(k+1)$-th layer in the re-optimized model \\
$\mathbf{b}_k^{(v)}$ & vector & bias vector of neurons in the $k$-th layer in the re-optimized model \\
$b_i^{(v)k}$ & real number & bias of the $i$-th neuron in the $k$-th layer in the re-optimized model \\
$\mathbf{W}_k^{(w)}$ & matrix & weight matrix from the $k$-th to the $(k+1)$-th layer in the further re-optimized model \\
$w_{ij}^{(w)k}$ & real number & weight from the $j$-th neuron in the $k$-th layer to the $i$-th neuron in the $(k+1)$-th layer in the further re-optimized model \\
$\mathbf{b}_k^{(w)}$ & vector & bias vector of neurons in the $k$-th layer in the further re-optimized model \\
$b_i^{(w)k}$ & real number & bias of the $i$-th neuron in the $k$-th layer in the further re-optimized model \\
$\mathbf{X}$ & vector & input vector \\
$\mathbf{H}$ & vector & hidden layer vector \\
$\mathbf{Y}$ & vector & output vector \\
$\mathbf{Y}_t$ & vector & output reference vector \\
$x_i$ & real number & the $i$-th element of the input vector \\
$h_i$ & real number & the $i$-th element of the hidden layer vector \\
$y_i$ & real number & the $i$-th element of the output vector \\
$y_{ti}$ & real number & the $i$-th element of the output reference vector \\
$t$ & time & time point of state occurrence \\
$\delta$ & time & time step \\
$L$ & real number & loss function value \\
$\nabla_{\mathbf{W}_1} L$ & real number & gradient of the loss function with respect to $\mathbf{W}_1$ \\
$\nabla_{\mathbf{b}_1} L$ & real number & gradient of the loss function with respect to $\mathbf{b}_1$ \\
$\nabla_{\mathbf{W}_2} L$ & real number & gradient of the loss function with respect to $\mathbf{W}_2$ \\
$\nabla_{\mathbf{b}_2} L$ & real number & gradient of the loss function with respect to $\mathbf{b}_2$ \\
$\eta$ & positive real number & learning rate \\
\end{longtable}}

{\renewcommand{\arraystretch}{1.2}
\setlength{\arrayrulewidth}{0.7pt}
\begin{longtable}{p{0.2\textwidth}p{0.25\textwidth}p{0.5\textwidth}}
\caption{Definition of Functions} \label{tab:functions} \\
\hline
Symbol & Number of Elements & Definition \\
\hline
\endfirsthead

\multicolumn{3}{c}{\tablename\ \thetable{} -- continued from previous page} \\
\hline
Symbol & Number of Elements & Definition \\
\hline
\endhead

\hline
\multicolumn{3}{r}{Continued on next page} \\
\endfoot

\hline
\endlastfoot

$\text{ReLU}(x)$ & 1 & rectified linear unit (ReLU) activation function \\
$\text{Identity}(x)$ & 1 & identity function \\
$\max$ & multivariate & maximum function \\
$\text{MSE}(\mathbf{Y},\mathbf{Y}_t)$ & 2 & mean squared error between vector $\mathbf{Y}$ and $\mathbf{Y}_t$ \\
$\text{ReLU}'(x)$ & 1 & derivative of $\text{ReLU}(x)$ \\
$\text{Target}(\mathbf{X})$ & 1 & objective function for input vector $\mathbf{X}$ \\
\end{longtable}}

{\renewcommand{\arraystretch}{1.2}
\setlength{\arrayrulewidth}{0.7pt}
\begin{longtable}{p{0.35\textwidth}p{0.6\textwidth}}
\caption{Definition of Predicates} \label{tab:predicates} \\
\hline
Symbol & Definition \\
\hline
\endfirsthead

\multicolumn{2}{c}{\tablename\ \thetable{} -- continued from previous page} \\
\hline
Symbol & Definition \\
\hline
\endhead

\hline
\multicolumn{2}{r}{Continued on next page} \\
\endfoot

\hline
\endlastfoot

$\text{NetworkStructure}^2(\text{layers},k)$ & model structure has $k$ layers \\
$\text{LayerSize}^2(k,j)$ & the $k$-th layer has $j$ neurons \\
$\text{Eq}^2(x,y)$ & $x$ equals $y$ \\
$\text{InputAt}^2(\mathbf{X},t)$ & input $\mathbf{X}$ at time $t$ \\
$\text{HiddenAt}^2(\mathbf{H},t)$ & hidden layer value is $\mathbf{H}$ at time $t$ \\
$\text{OutputAt}^2(\mathbf{Y},t)$ & output $\mathbf{Y}$ at time $t$ \\
$\text{InputAt}^3(\mathbf{X},\mathbf{Y}_t,t)$ & input vector $\mathbf{X}$ and label vector $\mathbf{Y}_t$ at time $t$ \\
$\text{Differentiable}^2(f,v)$ & function $f$ is differentiable with respect to $v$ \\
$\text{BelongsTo}^2(x,X)$ & element $x$ belongs to set $X$ \\
\end{longtable}}

\textbf{C.3 Formal Representations of Information at Each Stage of Model M}

Following MLT-MF, we build formal expressions for the following five stages:

\textbf{(1) Initial information of M}

According to the principles of OIT and based on MLT-MF, the model M is endowed with its initial information from the very beginning:
\begin{equation}
Io : S_{oo} (oo, To_h) \rightrightarrows S_M (M, To_m)
\end{equation}

Here, the ontology  $oo $ represents the designer's conceptual blueprint;  ${To}_{h} $ is the time at which the model is designed; and the ontological state  $S_{oo}(oo,{To}_{h}) $ is a set of WFFs that describe the initial operational logic of model M:
\begin{equation}
S_{oo}\left( oo,{To}_{h} \right) = \left\{ \varphi o_{1},\varphi o_{2},\varphi o_{3},\varphi o_{4},\varphi o_{5},\varphi o_{6},\varphi o_{7},\varphi o_{8},\varphi o_{9},\varphi o_{10},\varphi o_{11} \right\}
\end{equation}

The specific content of each formula is as follows:
\begin{itemize}
\item
  Network Structure Definition
\end{itemize}
\begin{equation}
\begin{aligned}
\varphi o_{1} = &\ \mathit{NetworkStructure}^{2}(layers,3) \land \\
& \mathit{LayerSize}^{2}(1,2) \land \mathit{LayerSize}^{2}(2,3) \land \\
& \mathit{LayerSize}^{2}(3,2)
\end{aligned}
\end{equation}
\begin{itemize}
\item
  Weight Initialization
\end{itemize}
\begin{equation}
\begin{aligned}
\varphi o_{2} = &\ \forall i \forall j \left( \mathit{BelongsTo}^{2}\left( i, \{1,2,3\} \right) \land \mathit{BelongsTo}^{2}\left( j, \{1,2\} \right) \right. \\
& \left. \rightarrow \exists v \left( \mathit{Eq}^{2}\left( w_{ij}^{(o)1}, v \right) \right) \right)
\end{aligned}
\end{equation}

\begin{equation}
\begin{aligned}
\varphi o_{3} = &\ \forall i \forall j \left( \mathit{BelongsTo}^{2}\left( i, \{1,2\} \right) \land \mathit{BelongsTo}^{2}\left( j, \{1,2,3\} \right) \right. \\
& \left. \rightarrow \exists v \left( \mathit{Eq}^{2}\left( w_{ij}^{(o)2}, v \right) \right) \right)
\end{aligned}
\end{equation}
\begin{itemize}
\item
  Bias Initialization
\end{itemize}
\begin{equation}
\varphi o_{4} = \forall i{BelongsTo}^{2}\left( i,\left\{ 1,2,3 \right\} \right) \rightarrow {\exists v(Eq}^{2}\left( b_{i}^{(o)1},v \right))
\end{equation}
\begin{equation}
\varphi o_{5} = \forall i{BelongsTo}^{2}\left( i,\left\{ 1,2 \right\} \right) \rightarrow {\exists v(Eq}^{2}\left( b_{i}^{(o)2},v \right))
\end{equation}
\begin{itemize}
\item
  Activation Function Definition
\end{itemize}
\begin{equation}
\varphi o_{6} = \forall x({Eq}^{2}(ReLU(x),max(0,x))) \land \forall x({Eq}^{2}(Identity(x),x))
\end{equation}
\begin{itemize}
\item
  Forward Propagation Rules
\end{itemize}
\begin{equation}
\begin{aligned}
\varphi o_{7} = &\ \forall X\forall t \left( \mathit{InputAt}^{2}(X,t) \land \mathit{BelongsTo}^{2}\left( X,\mathbb{R}^{2} \right) \rightarrow \right. \\
& \exists H \left( \mathit{BelongsTo}^{2}\left( H,\mathbb{R}^{3} \right) \land \mathit{HiddenAt}^{2}(H,t + \delta) \land \right. \\
& \left. \forall i \left( \mathit{BelongsTo}^{2}\left( i,\left\{ 1,2,3 \right\} \right) \rightarrow \right. \right. \\
& \left. \left. \mathit{Eq}^{2}\left( h_{i}, \mathrm{ReLU}\left( \sum_{j = 1}^{2}w_{ij}^{(o)1} \times x_{j} + b_{i}^{(o)1} \right) \right) \right) \right)
\end{aligned}
\end{equation}

\begin{equation}
\begin{aligned}
\varphi o_{8} = &\ \forall H\forall t \left( \mathit{HiddenAt}^{2}(H,t + \delta) \land \mathit{BelongsTo}^{2}\left( H,\mathbb{R}^{3} \right) \rightarrow \right. \\
& \exists Y \left( \mathit{BelongsTo}^{2}\left( Y,\mathbb{R}^{2} \right) \land \mathit{OutputAt}^{2}(Y,t + 2\delta) \land \right. \\
& \left. \forall i \left( \mathit{BelongsTo}^{2}\left( i,\left\{ 1,2 \right\} \right) \rightarrow \right. \right. \\
& \left. \left. \mathit{Eq}^{2}\left( y_{i}, \mathrm{Identity}\left( \sum_{j = 1}^{3}w_{ij}^{(o)2} \times h_{j} + b_{i}^{(o)2} \right) \right) \right) \right)
\end{aligned}
\end{equation}

\begin{itemize}
\item
  Backpropagation Rules
\end{itemize}
\begin{equation}
\begin{aligned}
\varphi o_{9} = &\ \forall X\forall Y_{t}\forall t \left( \mathit{InputAt}^{3}\left( X,Y_{t},t \right) \land \mathit{BelongsTo}^{2}\left( Y_{t},\mathbb{R}^{2} \right) \land \right. \\
& \left. \varphi o_{7} \land \varphi o_{8} \land \mathit{Eq}^{2}\left( L, \mathrm{MSE}\left( Y,Y_{t} \right) \right) \land \right. \\
& \left. \mathit{Eq}^{2}\left( \mathrm{MSE}\left( Y,Y_{t} \right), \frac{1}{2}\sum_{i = 1}^{2}\left( y_{i} - y_{ti} \right)^{2} \right) \right)
\end{aligned}
\end{equation}
\begin{align}
\varphi o_{10} &= \varphi o_{7} \land \varphi o_{8} \land \varphi o_{9} \nonumber\\
&\quad \land \mathit{Differentiable}^{2}(MSE(Y,Y_{t}),W_{1}^{(o)}) \nonumber\\
&\quad \land \mathit{Differentiable}^{2}(MSE(Y,Y_{t}),b_{1}^{(o)}) \nonumber\\
&\quad \land \mathit{Differentiable}^{2}(MSE(Y,Y_{t}),W_{2}^{(o)}) \nonumber\\
&\quad \land \mathit{Differentiable}^{2}(MSE(Y,Y_{t}),b_{2}^{(o)}) \nonumber\\
&\quad \land \mathit{Eq}^{2}(\nabla_{W_{1}^{(o)}}L, \nonumber\\
&\qquad ({(W_{2}^{(o)})}^{T}(Y - Y_{t})\scalebox{0.8}{$\bigodot$}\mathit{ReLU}'(W_{1}^{(o)}X + b_{1}^{(o)}))X^{T}) \nonumber\\
&\quad \land \mathit{Eq}^{2}(\nabla_{b_{1}^{(o)}}L, \nonumber\\
&\qquad {(W_{2}^{(o)})}^{T}(Y - Y_{t})\scalebox{0.8}{$\bigodot$}\mathit{ReLU}'(W_{1}^{(o)}X + b_{1}^{(o)})) \nonumber\\
&\quad \land \mathit{Eq}^{2}(\nabla_{W_{2}^{(o)}}L,(Y - Y_{t}){(\mathit{ReLU}(W_{1}^{(o)}X + b_{1}^{(o)}))}^{T}) \nonumber\\
&\quad \land \mathit{Eq}^{2}(\nabla_{b_{2}^{(o)}}L,(Y - Y_{t}))
\end{align}

\begin{itemize}
\item
  Learning Rules
\end{itemize}
\begin{equation}
\begin{split}
\varphi o_{11} = &\varphi o_{7} \land \varphi o_{8} \land \varphi o_{9} \land \varphi o_{10} \land {Eq}^{2}(W_{1}^{(u)},W_{1}^{(o)} - \eta\nabla_{W_{1}^{(o)}}L) \\
&\land {Eq}^{2}(b_{1}^{(u)},b_{1}^{(o)} - \eta\nabla_{b_{1}^{(o)}}L) \land {Eq}^{2}(W_{2}^{(u)},W_{2}^{(o)} - \eta\nabla_{W_{2}^{(o)}}L) \\
&\land {Eq}^{2}(b_{2}^{(u)},b_{2}^{(o)} - \eta\nabla_{b_{2}^{(o)}}L)
\end{split}
\end{equation}

In equation (A.14), the carrier M refers to the initially constructed model, and  ${To}_{m} $  denotes the time when model M is completed. Clearly, we have:
\begin{equation}
max({To}_{h}) \leq min({To}_{m})
\end{equation}

The state  $S_{M}(M,{To}_{m}) $ is a noisy-free enabling mapping of the state  $S_{oo}\left( oo,{To}_{h} \right) $ onto M, that is,
\begin{equation}
S_{M}\left( M,{To}_{m} \right) = \left\{ \psi o_{1},\psi o_{2},\psi o_{3},\psi o_{4},\psi o_{5},\psi o_{6},\psi o_{7},\psi o_{8},\psi o_{9},\psi o_{10},\psi o_{11} \right\}
\end{equation}
and
\begin{equation}
Io: \varphi o_i \rightrightarrows \psi o_i \quad (1 \leq i \leq 11)
\end{equation}

Here, the information  $Io $ enables the mapping from the initial ontological state of model M to its carrier state, that is, the initial state of M. In practice, this mapping is a complex process involving hardware implementation, software development, and hardware-software integration, inevitably requiring consumption of time and energy resources, and subject to various sources of noise. However, to simplify the discussion, we temporarily disregard these factors and explain the initial state of M using a completely ideal, and noise-free information model. The subsequent discussions on other types of information will follow this approach and will not be repeated.

\textbf{(2) Training Information of M}

Based on MLT-MF, the training information for model M is as follows:
\begin{equation}
It: S_{ot} (ot, Tt_h) \rightrightarrows S_{Mt} (Mt, Tt_m)
\end{equation}

Here, the ontology  $ot $ refers to the object described by the training information, and the occurrence time  ${Tt}_{h} $ denotes the point at which this object attains its corresponding state attributes. The ontological state  $S_{ot}(ot,{Tt}_{h}) $ represents the attributes of the object at time  ${Tt}_{h} $, which are described by a set of WFFs:
\begin{equation}
S_{ot}\left( ot,{Tt}_{h} \right) = \left\{ \varphi t_{1},\varphi t_{2} \right\}
\end{equation}
\begin{itemize}
\item
  Vector Definition
\end{itemize}
\begin{equation}
\varphi t_{1} = \forall X\forall Y_{t}({BelongsTo}^{2}(X,\mathbb{R}^{2}) \land {BelongsTo}^{2}(Y_{t},\mathbb{R}^{2}))
\end{equation}

\begin{itemize}
\item
  Label Consistency
\end{itemize}
\begin{equation}
\varphi t_{2} = \forall X\forall Y_{t}{Eq}^{2}(Target(X),Y_{t})
\end{equation}

The carrier  $Mt $   serves as the interface device through which model M receives training data, and  ${Tt}_{m} $  is the time at which model M is trained. The carrier state  $S_{Mt}(Mt,{Tt}_{m}) $ is then described by the set of WFFs:
\begin{equation}
It: \varphi t_i \rightrightarrows \psi t_i \quad (1 \leq i \leq 2)
\end{equation}

\textbf{(3) Optimized Information After M Has Learned from Training}

Based on MLT-MF, after learning the training information $It $, model M updates its initial information  $Io $ to a new, optimized information state:
\begin{equation}
Iu: S_{ou} (ou, Tu_h) \rightrightarrows S_M (M, Tu_m)
\end{equation}

Here, the ontology  $ou $ is a fusion of M and the object described by the training information; the occurrence time  ${Tu}_{h} $ is the moment M learns from the training data; and the ontological state  $S_{ou}(ou,{Tu}_{h}) $ represents the theoretical state of M after training:
\begin{equation}
S_{ou}\left( ou,{Tu}_{h} \right) = \left\{ \varphi u_{1},\varphi u_{6},\varphi u_{7},\varphi u_{8},\varphi u_{9},\varphi u_{10},\varphi u_{11} \right\}
\end{equation}

Since the structure and algorithms of M remain unchanged before and after training,  $\varphi u_{1},\varphi u_{6} $, and  $\varphi u_{9} $ are exactly the same as  $\varphi o_{1},\varphi o_{6} $, and  $\varphi o_{9} $ respectively. Moreover, as the weights and biases after training are already assigned values in the original information (A.25), there is no need to include equivalent formulas for  $\varphi o_{2},\varphi o_{3},\varphi o_{4} $, or  $\varphi o_{5} $  in  $S_{ou}\left( ou,{Tu}_{h} \right) $. For the remaining formulas  $\varphi u_{7},\varphi u_{8},\varphi u_{10} $, and  $\varphi u_{11} $, it suffices to systematically replace  $W_{k}^{(o)} $,  $w_{ij}^{(o)k} $,  $b_{k}^{(o)} $,  $b_{i}^{(o)k} $,  $W_{k}^{(u)} $, $w_{ij}^{(u)k} $,  $b_{k}^{(u)} $ and  $b_{i}^{(u)k} $ in  $\varphi o_{7},\varphi o_{8},\varphi o_{10} $, and  $\varphi o_{11} $  with the corresponding symbols  $W_{k}^{(u)} $,  $w_{ij}^{(u)k} $,  $b_{k}^{(u)} $,  $b_{i}^{(u)k} $,  $W_{k}^{(v)} $,  $w_{ij}^{(v)k} $,  $b_{k}^{(v)} $ and  $b_{i}^{(v)k} $ in  $S_{ou}\left( ou,{Tu}_{h} \right) $.

Similarly, the carrier state
\begin{equation}
S_{M}\left( M,{Tu}_{m} \right) = \left\{ \psi u_{1},\psi u_{6},\psi u_{7},\psi u_{8},\psi u_{9},\psi u_{10},\psi u_{11} \right\}
\end{equation}

is a noisy-free enabling mapping from  $S_{ou}\left( ou,{Tu}_{h} \right) $ onto M, representing the state of M after completing its training.

\textbf{(4) Input Information for M}

Processing input information is a fundamental requirement for any model. Based on the MLT-MF framework, the input information for M is defined as:
\begin{equation}
Iq: S_{oq} (oq, Tq_h) \rightrightarrows S_{Mq} (Mq, Tq_m)
\end{equation}

Here,  $oq $   denotes the object described by the input information, and  ${Tq}_{h} $ is the time at which the object\textquotesingle s state occurs. The state  $S_{oq}(oq,{Tq}_{h}) $ can take two possible forms:

\begin{itemize}
\item
  The first scenario involves only input information being processed by  M, in which case
\end{itemize}
\begin{equation}
S_{oq}(oq,{Tq}_{h}) = \left\{ \varphi q_{1} \right\}
\end{equation}
\begin{equation}
\varphi q_{1} = \forall X{BelongsTo}^{2}(X,\mathbb{R}^{2})
\end{equation}

\begin{itemize}
\item
  The second scenario considers online learning, where
\end{itemize}
\begin{equation}
S_{oq}(oq,{Tq}_{h}) = \left\{ \varphi q_{1},\varphi q_{2} \right\}
\end{equation}
\begin{equation}
\varphi q_{1} = \forall X\forall Y_{t}({BelongsTo}^{2}(X,\mathbb{R}^{2}) \land {BelongsTo}^{2}(Y_{t},\mathbb{R}^{2}))
\end{equation}
\begin{equation}
\varphi q_{2} = \forall X\forall Y_{t}{Eq}^{2}(Target(X),Y_{t})
\end{equation}

The carrier  $Mq $   for  $Iq $ serves as the input device for M, which may or may not be the same as the carrier  $Mt $  for the training information  $It $. The state  $S_{Mq}(Mq,{Tq}_{m}) $ is a noisy-free enabling mapping from  $S_{oq}(oq,{Tq}_{h}) $ onto  $Mq $.

\textbf{(5) Information After M Processes Input, Produces Output, and Undergoes Further Optimization}

There are two scenarios for M after it processes input and produces output: one corresponds to the first case in (4), where only output is produced without further optimization of the model itself; the other corresponds to the second case in (4), where, in the context of online learning, the model both produces output and undergoes further self-optimization. Since the latter is more comprehensive, we will focus on this scenario below.

In this case, the output information $Ir: S_{or} (or, Tr_h) \rightrightarrows S_{Mr} (Mr, Tr_m) $ has an ontological state defined as:
\begin{equation}
S_{or}\left( or,{Tr}_{h} \right) = \left\{ \varphi r_{1} \right\} = \left\{ {\varphi o_{7} \land \varphi o_{8} \land OutputAt}^{2}(Y,{Tr}_{h}) \right\}
\end{equation}

The carrier state  $S_{Mr}(Mr,{Tr}_{m}) $ is a noisy-free enabling mapping from  $S_{or}\left( or,{Tr}_{h} \right) $ to the output device of M.

On the other hand, since we are considering the scenario where M performs online learning, the model now carries further optimized information:
\begin{equation}
Iv: S_{ov} (ov, Tv_h) \rightrightarrows S_{M} (M, Tr_m)
\end{equation}

This structure is almost identical to the optimized information  $Iu $ after M has learned from training. The ontological state of $Iv $  is:
\begin{equation}
S_{ov}\left( ov,{Tv}_{h} \right) = \left\{ \varphi v_{1},\varphi v_{6},\varphi v_{7},\varphi v_{8},\varphi v_{9},\varphi v_{10},\varphi v_{11} \right\}
\end{equation}

Here, $\varphi v_{1}$, $\varphi v_{6}$, and $\varphi v_{9}$ are identical to $\varphi u_{1}$, $\varphi u_{6}$, and $\varphi u_{9}$ respectively. For the remaining formulas, it suffices to systematically replace the parameters 
\begin{align}
&W_{k}^{(u)}, w_{ij}^{(u)k}, b_{k}^{(u)}, b_{i}^{(u)k}, W_{k}^{(v)}, w_{ij}^{(v)k}, b_{k}^{(v)}, b_{i}^{(v)k} \nonumber
\end{align}
in $S_{ou}(ou,{Tu}_{h})$ with the corresponding parameters 
\begin{align}
&W_{k}^{(v)}, w_{ij}^{(v)k}, b_{k}^{(v)}, b_{i}^{(v)k}, W_{k}^{(w)}, w_{ij}^{(w)k}, b_{k}^{(w)}, b_{i}^{(w)k} \nonumber
\end{align}
to obtain the updated formulas, where $W_{k}^{(w)}$, $w_{ij}^{(w)k}$, $b_{k}^{(w)}$, and $b_{i}^{(w)k}$ are the weights and biases of M after further optimization.

Similarly, the carrier state
\begin{equation}
S_{M}\left( M,{Tv}_{m} \right) = \left\{ \psi v_{1},\psi v_{6},\psi v_{7},\psi v_{8},\psi v_{9},{\psi v}_{10},\psi v_{11} \right\}
\end{equation}
is a noisy-free enabling mapping from $S_{ov}\left( ov,{Tv}_{h} \right) $ onto M.

At this point, based on MLT-MF, we have constructed the formal well-formed formula representations for all ontological and carrier states of each component of the feedforward neural network model M, as illustrated in Fig. 7. Although M is a simple model, it contains all the essential elements of a neural network, and the discussion has not relied on any special-case treatment of M. This demonstrates that MLT-MF is capable of supporting the construction of WFF representations for a wide variety of neural networks, providing a solid foundation for detailed analysis and optimization.

\end{document}